# Food grade nanoemulsions preparation by rotor-stator homogenization

Dilek Gazolu-Rusanova[1], Ivan Lesov[1],

Slavka Tcholakova[1*], Nikolai Denkov[1], Badreddine Ahtchi[2]

[1]*Department of Chemical and Pharmaceutical Engineering, Faculty of Chemistry and Pharmacy, Sofia University, 1164 Sofia, Bulgaria*
[2]*PepsiCo Global R&D, 3 Skyline Drive, Hawthorne, NY 10532, USA*

***Corresponding author**:
Prof. Slavka Tcholakova
Department of Chemical and Pharmaceutical Engineering
Faculty of Chemistry and Pharmacy, Sofia University
1 James Bourchier Ave., 1164 Sofia
Bulgaria

Phone: (+359-2) 962 5310
Fax:    (+359-2) 962 5643
E-mail: SC@LCPE.UNI-SOFIA.BG

# ABSTRACT

High-pressure homogenizers, typically used for producing nanoemulsions at the industrial scale, are energy and maintenance intensive, and limited to produce only dilute, low viscosity nanoemulsions. We propose an alternative approach to produce dilute to concentrated food-grade nanoemulsions with droplet size ranging between 100 and 500 nm using rotor-stator homogenization. Gum Arabic (GA) or modified starch (MS) was used as both viscosity modifier and emulsion stabilizer. GA and MS have relatively low surface activity compared to the common low-molecular-mass surfactants used typically for nanoemulsion preparation. The main differences between GA and MS are the lower viscosity of the GA solutions, compared to MS solutions, and the faster adsorption of MS, as compared to GA. The obtained results show that stable nanoemulsions are formed by rotor-stator homogenization when the rapidly adsorbing MS is used as emulsifier. Much larger drops are formed during emulsification with GA, which is due to significant drop-drop coalescence in the respective emulsions. The experimental results for the nanoemulsions prepared with MS are well-described by the theoretical expression for emulsification in turbulent viscous regime, after proper account for the effects of temperature and drop-drop interactions in the sheared emulsions.

# 1. Introduction

Emulsions are dispersions of immiscible liquid phases, where one of the liquids is distributed among the other in the form of small drops. Based on the drop size, emulsions are typically presented as 'conventional' or macroemulsions (drop diameters larger than 1 μm), nanoemulsions (with droplet size typically between 20 nm and 500 nm according to Salem & Ezzat 2019) and microemulsions (drop diameters in the range between 10 and 50 nm).

One of the most challenging problems in the application of nanoemulsions is their preparation, as one should apply a certain strategy for minimization of the droplet sizes and, therefore, these emulsions are prepared usually by high-energy homogenizers (McClements & Jafari 2018, Jamali, Assadpour & Jafari 2019). The formation, stability, and properties of nanoemulsions often depend on the physicochemical properties of the oil phase, e.g., its polarity, water-solubility, interfacial tension, refractive index, viscosity, density, phase behavior, and chemical stability (Anton, Gayet, Benoit, & Saulnier, 2007; Anton & Vandamme, 2009; McClements, 2005; Tadros et al., 2004; Wooster et al., 2008; McClements, 2011; McClements & Rao, 2011; Jafari, He & Bhandari 2007; Jafari, Beheshti & Assadpour 2013). For example, the formation of nanoemulsions using medium- or long-chain triglyceride oils (as the oils used in the current study) is often challenging due to their relatively low polarity, high interfacial tension, and high viscosity. However, once a nanoemulsion has been created using triglyceride oils (TGO), it is often rather stable, especially regarding Ostwald ripening, due to the low molecular solubility of these oils (Piorkowski & McClements, 2013; Shamsara, Jafari & Muhidinov 2017).

The composition of the aqueous phase affects important physicochemical parameters, such as the interfacial tension, viscosity and mass density of the continuous phase, which all play a key role for the drop breakup during emulsification. The selection of appropriate emulsifiers is one of the key elements for successful emulsification and affects the long-term stability of the emulsions obtained. In the high-energy methods for nanoemulsion formation, the emulsifier facilitates droplet breakup by lowering the interfacial tension and stabilizes the formed drops against coalescence. In the low-energy approaches, the emulsifiers facilitate the spontaneous formation of small droplets by producing ultra-low interfacial tension under certain conditions, e.g. upon temperature variation and at high surfactant concentrations. Very often, the formation of nanoemulsions using low-energy methods requires the addition of co-surfactants or co-solvents (Flanagan & Singh, 2006; Yaghmur, Aserin & Garti, 2002).

The high-pressure homogenizers are the most widely used devices for the production of nanoemulsions. The mechanical breakup of the oil phase into tiny droplets is realized as a result of the intense disruptive inertial and cavitation forces in these homogenizers (McClements & Rao, 2011). Recently, there is a growing interest in nanoemulsion formation using rotor-stator homogenizers. These devices have several advantages compared to the other techniques for emulsion preparation, such as: relatively easy to install and work with;

require comparably low costs of investment; relatively high throughput; compatible with viscous systems; allow large volume emulsion preparation, etc. Here the effective drop breakup energy originates from the forces of inertia and shearing in turbulent flow (Urban, Wagner, Schaffner, Röglin & Ulrich, 2006a). However, decreasing the drop size below 1 µm in the rotor-stator devices is very difficult (Jafari, He & Bhandari, 2006, 2007a, 2007b).

The most often used strategies for reducing the droplet size in rotor-stator homogenizers are: increasing the rotor speed, decreasing the gap size, changing the rotor design, increasing the residence time of the droplets in the disruption zone (i.e. long emulsification time). Nevertheless, manipulating all these process parameters could be insufficient to produce nanoemulsions if the formulation composition is not appropriate. To achieve effective drop breakup, emulsifiers with fast adsorption kinetics and very low interfacial tension are preferred (Van der Schaaf & Karbstein, 2018). There are several authors that were successful in the preparation of nanoemulsions using rotor-stator devices (El-Jaby, McKenna & Cunningham, 2007; El-Jaby, Cunningham & McKenna, 2009; Han et al., 2012; Karthik & Anandharamakrishnan, 2016; Scholz & Keck, 2015; Wolf, Koehler & Schuchmann, 2013; Shamsara, Jafari & Muhidinov 2017). Thus, McKenna & Cunningham, 2007 prepared nanoemulsions using low viscosity hexadecane in the presence of fast adsorbing sodium dodecylbezene sulfate (SDBS) of high concentration in the aqueous phase. Han et al. 2012 prepared nanoemulsions by combing rotor-stator homogenization and ultrasound. Karthik & Anandharamakrishnan 2016 prepared nanoemulsions by rotor-stator homogenizer but they showed that much smaller droplets are formed when high pressure homogenizer is used for the same emulsion composition. Scholz & Keck, 2015 prepared diluted 5 % nanoemulsions from Miglyol 812 using a mixture of Tween 80 and Span 80 at high surfactant concentration, high mixing rate of 36000 rpm, and long emulsification time of 5 min. Wolf, Koehler & Schuchmann 2013 demonstrated that water-in-oil nanoemulsions can be prepared by Colloidal Mill if PGPR is used as emulsifier. From this brief summary one sees that all these studies used low-molecular-mass surfactants for nanodroplets creation and stabilization.

Gum Arabic (GA) and modified starch (MS), used in the current study, do not fulfill the requirements for low interfacial tension, which makes the task of nanoemulsion formation more challenging. Gum Arabic is primarily extracted from acacia trees, *Acacia Senegal* and *Acacia Seyal* (Montenegro, Boiero, Valle & Borsarelli, 2012) and has complex chemical structure. In general, the GA molecules consist of a hydrophilic polysaccharide groups (1,3-linked α-galactopyranose monomers and 1,6-linked galactopyranose side chains) which contribute to around 88% of the total GA mass. They are attached to polypeptide backbone – arabinogalactan protein complex and glycoprotein (Al-Assaf, Phillips & Williams, 2005; Dickinson, 1992, 2003; Garti & Leser, 2001; Idris, Williams & Phillips, 1998; Islam, Phillips, Sljivo, Snowden & Williams, 1997; Randall, Phillips & Williams, 1988; Renard, Lavenant-Gourgeon, Ralet & Sanchez, 2006). The arabinogalactan protein complex is around 11 wt% and around 10-12 wt% of it is the protein content. Another 1 wt % is the glycoprotein. The two proteins, which are around 2.4 wt% from the total GA mass, are considered to be the surface-active part in the aforementioned structure. The protein content is very low with respect to the total molecule content and Gum Arabic has a relatively low affinity to oil-water interfaces, compared to most other surface-active biopolymers. Therefore, relatively high concentrations of Gum Arabic are required to form stable emulsions with small droplet size. However, the high concentration of GA often leads to undesired depletion flocculation between the neighboring drops (Chanamai & McClements, 2001).

Natural starches have poor surface activity due to their hydrophilic nature. Therefore, to make them adsorb at an oil-water interface, a hydrophobic modification is made. One of the most commonly used modifications is the addition of octenyl succinate. This hydrophobic group anchors the starch molecules to the oil droplet surface, while the hydrophilic starch chains protrude into the aqueous phase and protect the droplets against aggregation through steric repulsion. Similarly to GA, the modified starch (MS) has a relatively low interfacial activity when compared to conventional surfactants, but often adsorbs faster than Gum Arabic (Erni et al., 2007). This faster adsorption results in better performance of the modified starch as emulsifier and in the formation of smaller droplets (Chanamai & McClements, 2002).

The major aim of the present study is to find the conditions at which stable emulsions with droplet sizes in the range between 100 and 500 nm could be obtained for short emulsification time via rotor-stator homogenization. Gum Arabic (GA) and modified starch (MS) were tested as both emulsifiers and viscosity thickeners, and two types of triglyceride oils were compared as dispersed phase. The main advantage of the proposed procedure for formation of nanoemulsions is the possibility to use higher concentration of oil as compared to high pressure homogenization, without adding low-molecular-mass surfactants for droplet stabilization.

## 2. Materials and methods.

### 2.1. Materials.

The studied emulsifiers are modified starch (Starch sodium octenyl succinate, EmCap$^{TM}$ Starches, Cargill$^{TM}$) and Gum Arabic (TIC Pretested® GumArabic Spray Dry Powder, TIC Gums). The concentration of the emulsifiers in the aqueous phase was varied between 15.6 and 35 wt%. All solutions contained sodium benzoate (Sigma) and citric acid (Sigma). The mass ratios of benzoate/emulsifier and citric acid/emulsifier were kept constant – $7.1\times10^{-3}$ and $1.1\times10^{-3}$, respectively. Deionized water purified by Milli-Q Organex system (Millipore, USA) was used for solution preparation. The preparation of the aqueous phase includes a hydration step, where the homogenized mixture of emulsifiers, preservatives and water is transferred into a water bath, set to certain temperature depending on the emulsifier used. The temperature and the period of hydration are 5 min at 60 °C for MS and 1 hour at 50 °C for GA. During this hydration step, the solutions were periodically stirred by spoon. After the hydration period, the solution of MS was used while it was still hot and the solution of GA was cooled down to 35 °C and then used for emulsion preparation.

As a dispersed phase, we used Triglyceride Oil 36 (TGO-36) and Triglyceride Oil 19 (TGO-19). The viscosities of these oils, $\eta_O$, at 30 °C are 36 mPa.s for TGO-36 and 19 mPa.s for TGO-19. Increasing the temperature from 30 to 65 °C leads to *ca.* two times decrease of $\eta_O$. The oil viscosity is constant and does not depend on the shear rate in the measured range 1-500 s$^{-1}$. Two main series of experiments were performed depending on the oil concentration – emulsions with fixed oil concentration at 17.9 wt%, and emulsions in which the oil concentration was varied between 16 and 25 wt%. Note that all emulsions prepared with HPH were with fixed emulsifier-oil ratio: the oil concentration in the emulsions was fixed at 10 wt% and the emulsifier concentration in the aqueous phase was 15.6 wt%. All materials were used as received.

### 2.2. Procedures for emulsion preparation.

To prepare the coarse emulsions, during constant hand stirring with a spoon, the necessary amount of oil was slowly added into the emulsifier solution for 120 s and then stirred for additional 120 s. Afterwards, the coarse emulsions were homogenized either with IKA MagicLAB apparatus, equipped with ultra turrax module (UTL) with Generator 6F, or with High-pressure homogenizer (HPH, GEA Niro Soavi PandaPLUS 2000). The MagicLab was connected to a pump (ISMATEC; MCP-CPF Process IP65; Pump head - FMI212/QP.Q2.CSC/9004) in order to facilitate the emulsion flow through the UTL homogenizer. The emulsions were passed 3 times through the homogenizer. For UTL, we varied the rotor speed from 15 000 rpm to 25 000 rpm. The pump operated at rotation rates from 300 to 900 rpm depending on emulsification conditions (viscosity, rotor speed, temperature increase). The pressure in HPH was fixed to 5000 psi (34.47 MPa). Note that during emulsification, the emulsions in the funnel of the HPH and MagicLAB were constantly stirred to suppress drop creaming and to make sure that relatively homogeneous emulsion passes through the homogenizer. The used rotor type (generator 6F) has a complex geometry with three concentric high-shear gaps. During operation, the emulsion enters first the inner gap and passes toward the outer gaps. The dimensions of the gap with the largest diameter, where the highest shear rate is realized, are the following: radius 15 mm and gap-width 200 µm. The temperature of the emulsions after passing through the homogenizer was measured after each pass. The obtained $T_{OUT}$ of the emulsions are presented in Section C in Supplementary material.

## 2.3. Determination of the drop size distribution in the emulsions.

The drop size in the emulsions was determined by Dynamic Light Scattering (DLS) using Zetasizer Nano ZS instrument (Malvern Instruments), at a temperature of 25 °C, scattering angle 173°, and laser wavelength of 633 nm. The samples for DLS analysis were taken after gentle homogenization of the obtained emulsion at the respective conditions of interest (number of passes, rpm, etc.). The samples were diluted in 1 wt % solution of sodium dodecyl sulfate (SDS) or sodium lauryl ether sulfate (SLES) until diluted emulsion with relatively low opacity was obtained. The viscosity of this diluted emulsion was assumed to be equal to that of the medium (water). At least 3 measurements were carried out for each emulsion to check the reproducibility. We used two mean volume diameters, $d_{VM}$ and $d_{V50}$, and the volume-95% diameter, $d_{V95}$, as characteristics of the drop size distribution. $d_{VM}$ is calculated by multiplying the mode of the sub-peak in the obtained drop size histogram by its area in %. When multiple peaks appear in the size distribution, $d_{VM}$ is equal to the sum of the product mode×area of each peak. $d_{V50}$ and $d_{V95}$ are defined as the diameters, for which 50% and 95% by volume of the dispersed oil is contained in drops with $d \leq d_{V50}$ or $d_{V95}$, respectively. These diameters were determined from the measured cumulative size-distribution histograms of the drops in the emulsions. The diameter $d_{V95}$ was used as an experimentally accessible measure of the maximum droplet size. The polydispersity of the emulsions prepared with UTL was determined in terms of the Polydispersity Index (PDI) which is defined as the square of the ratio between the width of the distribution and the mean diameter, viz. the ratio $d_{V84}/d_{V50}$.

## 2.4. Rheological properties of the aqueous phases and of the formed emulsions.

The rheological properties of the oils, aqueous phases, and some of the emulsions were characterized by a rotational rheometer Bohlin Gemini (Malvern Instruments, UK). We employed a cone (2° angle) and plate geometry, with 60 mm cone diameter for the oils and solutions, and 40 mm cone diameter for the emulsions. The rheological response of the solutions, oils and emulsions was measured at steady shear experiments at 30 °C (for oils) and 65 °C (for oils, aqueous phases and emulsions). Before each type of rheological test, the samples were left to thermally equilibrate in the rheometer for 120 s at the respective temperature. The shear rate was varied stepwise (3 s integration time and 2 s delay after each steps) in the range 0.1 ÷ 500 s$^{-1}$ for the experiments with oils and emulsifier solutions, and in the range 0.1 ÷ 1000 s$^{-1}$ for the experiments with emulsions. For the experiments with the aqueous phases, pre-shear at 100 s$^{-1}$ shear rate and 65 °C for 120 s was applied prior to the steady shear experiments. Additionally, the viscosity of the oils and aqueous phases was measured as a function of temperature, at constant shear rate, (100 s$^{-1}$) by gradually increasing the temperature from 30 to 75 °C (rate 3 °C/min). In these experiments, pre-shear at 100 s$^{-1}$ shear rate and 30 °C for 120 s was applied prior to the measurement of the aqueous phases.

**2.5. Measurement of interfacial tension.**

The interfacial tension of the oil-water interface, $\sigma_{OW}$, was measured via drop shape analysis on a DSA100m (Krüss GmbH, Hamburg, Germany), equipped with a thermostatic chamber (± 0.1 °C). The profile of a drop of aqueous solution in bulk oil (pendant drop) was fitted by Laplace equation of capillarity, and $\sigma_{OW}$ was determined as a free parameter. The interfacial tension was measured mainly at 65 ± 1 °C. For some systems, the measurements were performed at temperatures between 30 and 80 °C to check for the effect of temperature. The duration of each measurement was set to 30 min. This method is suitable for measuring the equilibrium interfacial tension and adsorption kinetics after 1 s, which is much longer time scale when compared to the characteristic time for drop breakage. On the other hand, this time scale is comparable to the characteristic time of one pass of the emulsion through the equipment, which is important when considering the effect of droplet-droplet coalescence.

The mass density of the aqueous solutions and oils was measured with DMA35 Portable density meter (Anton Paar, Austria) at 25, 30, 35, 40 and 45 ± 0.1 °C. The accuracy of the measurement was ± 0.001 g/cm$^3$. From the best fit to the data, we determined the linear equations describing the temperature dependence of the mass density for each system. These equations were used to calculate the mass density of the aqueous solutions and oils at the desired temperature. The parameters from the linear fit of the data are presented in Table A.1-1in Supplementary material.

**3. Experimental results**

**3.1. Viscosity and interfacial properties of the studied aqueous phases.**

Here, we present the comparison of the viscosity and interfacial properties of the studied aqueous phases of MS and GA. The information collected from these experiments is used in the interpretation of the results obtained in the emulsification experiments. We measured the viscosity of MS and GA solutions with different concentrations of the respective biopolymer, as a function of the shear rate and temperature (results shown in Figure A.1 in Supplementary material). As expected, the viscosity increases with the increase of biopolymer concentration. All studied solutions have nearly Newtonian behavior – the viscosity depends weakly on the shear rate. The only exception is the solution of 35 wt % GA, which has shear thinning behavior at lower shear rates, but reaches plateau above 10 s$^{-1}$. Figure 1 shows the direct comparison of the viscosities of MS and GA solutions. At the same concentration, the viscosity of MS solutions is higher than that of GA solutions. The only exception is the lowest studied concentration, 15.6 wt%, where the viscosities of the two biopolymer solutions are similar. In most of the experiments we used solutions of MS and GA of 30.4 wt%. The viscosity of the respective MS solution (110 mPa.s at 65 °C) is almost twice higher than the viscosity of GA solution (59 mPa.s at 65 °C).

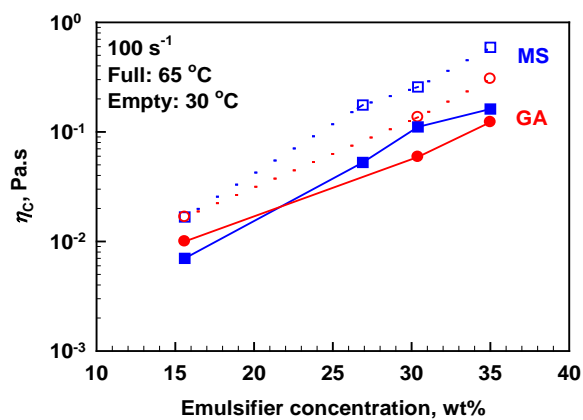

**Figure 1**. Viscosity of the solutions of MS (blue squares) and GA (red circles) as a function of emulsifier concentration, measured at 100 s$^{-1}$ and at 30 °C (empty symbols) and 65 °C (full symbols).

The interfacial tension of the studied solutions at the TGO-36-water interface, $\sigma_{OW}$, measured at different times after the interface formation, is presented as a function of the biopolymer concentration in Figure 2A and of temperature in Figure 2B. The kinetics of the interfacial tension decrease is shown in Figures A.2 in Supplementary material. One sees no significant change in the interfacial tension for both emulsifiers studied when the emulsifier concentration is increased from 15.6 wt% up to 35.0 wt%. The effect of temperature on $\sigma_{OW}$ is also relatively weak for both emulsifiers. Note, that the comparison in Figure 2B is made with solutions of MS and GA which have similar viscosity and, therefore, the concentrations of the two emulsifiers are different in these measurements. However, there is a significant difference in $\sigma_{OW}$ kinetics – MS adsorbs much faster than GA. The tension $\sigma_{OW}$ of GA solutions after formation of the drop is around 20 mN/m and it decreases gradually to around 11 mN/m for 30 min, whereas the interfacial tension of the MS solutions starts of ≈ 15 mN/m and decreases down to ≈ 13 mN/m for 30 min. The faster adsorption of MS, as compared to GA, is related to the fact that surface active species in MS have molecular mass < 10$^5$ g/mol according to the MS producer, whereas the surface active protein in GA has molecular mass of 2.5×10$^5$ g/mol (Renard et al. 2006) and is of relatively low concentration. One sees also that the interfacial tensions are considerably higher for both emulsifiers when compared to the interfacial tensions of the common low-molecular-mass surfactants used in nanoemulsion formation.

To compare the type of emulsion stabilization by the MS and GA emulsifiers, we performed microscope observations of emulsion films in capillary cell (Gazolu-Rusanova et al. 2020). Figure 2C presents video images of emulsion films of type oil-water-oil, formed from aqueous solutions of MS and GA. One sees a significant difference in the emulsion film behavior for the two emulsifiers. MS stabilizes thinner and homogeneous emulsion films, whereas the films stabilized by GA have irregular thickness (i.e. the films contain trapped protein aggregates) and remain inhomogeneous in thickness even 30 min after film formation.

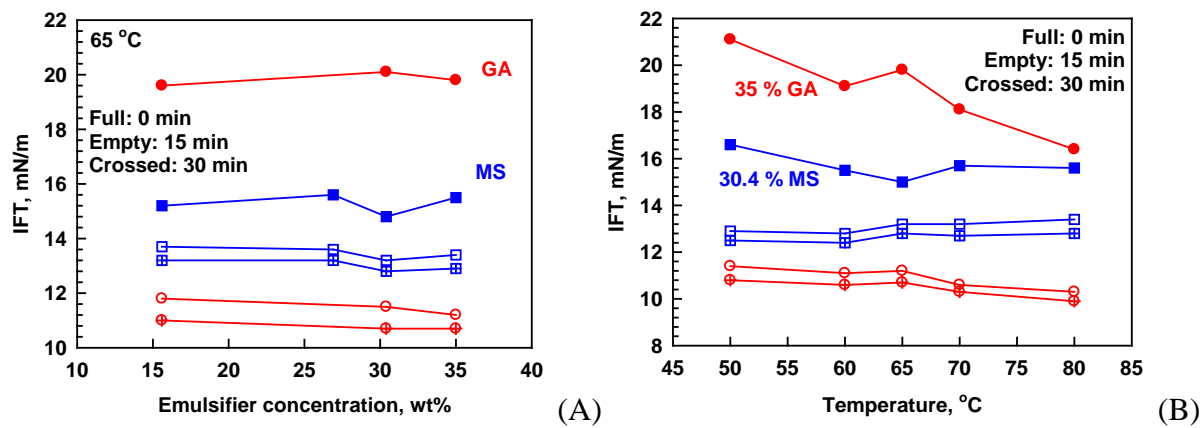

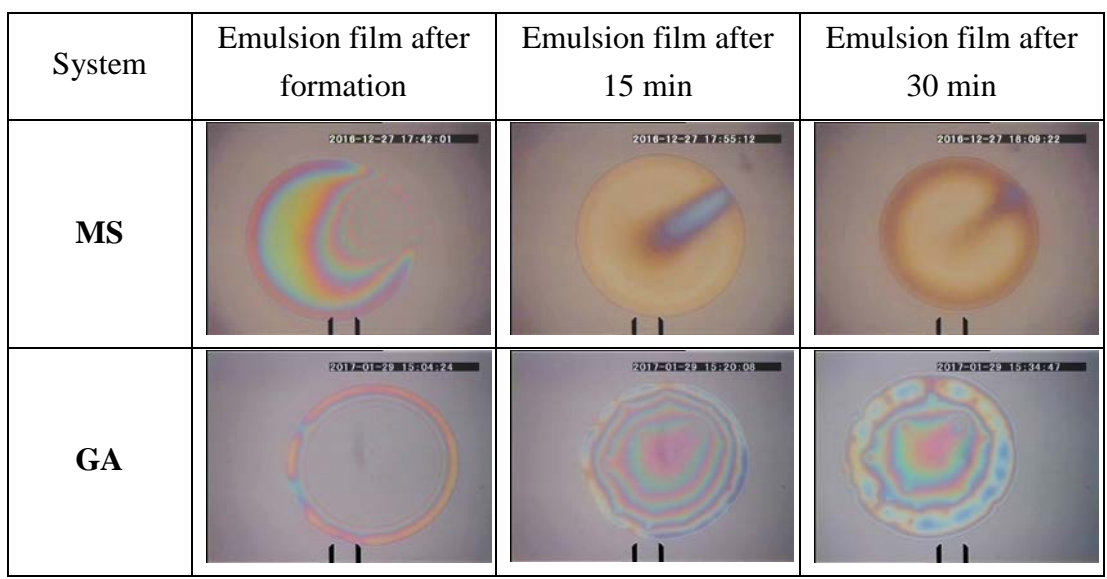

**Figure 2. (A, B)** TGO-36-water interfacial tension 0 min (full symbols), 15 min (empty symbols) and 30 min (crossed symbols) after formation of the pendant drop: (A) as a function of emulsifier concentration as measured at 65 °C, and (B) as a function of temperature. Concentrations of the biopolymers in wt% are color coded in the figure. The interfacial tension of TGO-36-water interface without added emulsifier is 24.5 mN/m at 65 °C; (C) Emulsion films formed in capillary cell and observed under optical microscope in reflected white light. Concentration of the emulsifiers: 30.4 wt% MS and 35 wt% GA. Distance between the vertical bars, 50 μm.

**3.2. Formation of TGO-36 nanoemulsions with UTL.**

To study the ability of MS and GA to stabilize nanoemulsions, we performed emulsification experiments with HPH at 5000 psi (34.47 MPa) and UTL homogenizer at three rotor speeds: 15000, 20000 and 25000 rpm. Since the experiments on HPH do not allow us to work at high viscosities, we kept the concentration of the emulsifier and the oil relatively low in the emulsions prepared with this homogenizer – 15.6 wt% emulsifier in the aqueous phase and 10 wt% TGO-36 in the final emulsion. For the experiments with UTL, we fixed the concentration of TGO-36 to 17.9 wt% and the concentration of MS and GA in the aqueous solutions to 30.4 wt%. The emulsifier/oil ratio is 1.4 for the experimental series on both homogenizers. The performance of each emulsifier is presented separately and then their efficiency for the two homogenizers is compared.

The mean, $d_{VM}$, and maximum droplet diameters, $d_{V95}$, obtained from emulsions stabilized with MS, are shown in Figure 3 ($d_{V50}$ and droplet size distributions are presented in Figure B.1 in Supplementary material). Both $d_{VM}$ and $d_{V95}$ decrease with the increase of the rotor speed and the number of passes through the equipment. When the emulsification is performed at 25 krpm rotor speeds, the characteristics of the obtained emulsions are very similar to those of the emulsions prepared by HPH. The polydispersity of the emulsions obtained with UTL at 20 and 25 krpm is lower when compared to the polydispersity of the emulsions prepared with HPH ($d_{V84}/d_{V50} \approx 1.9$ for emulsions prepared with HPH and $\approx 1.6$ for emulsions prepared with UTL). These important results demonstrate that oil-in-water nanoemulsions could be formed in a rotor-stator homogenizer not only using rapidly adsorbing emulsifiers with very low $\sigma_{OW}$, but also with relatively slowly adsorbing polymers which have significantly higher $\sigma_{OW}$ compared to Tween surfactants for example.

The results obtained in this series of experiments prove that the rotor-stator homogenizer UTL-MagicLab can produce emulsions with droplet sizes as small as in the emulsions obtained with HPH, at similar emulsifier/oil ratio. Similar results for comparable drop sizes in emulsion prepared by HPH and ART MICRRA D27 rotor-stator system were reported previously by Scholz & Keck 2015 for 5 % emulsion of medium chain triglycerides Miglyol 812 in water, stabilized by the surfactant mixture 5 wt % Tween 80+Span 80.

We performed emulsification experiments also with GA and the obtained results are shown in Figure 3C,D. One sees that for this emulsifier we cannot obtain nanoemulsions, when UTL is used. Unfortunately, we could not increase the rotor speed further, since the UTL module was getting jammed at high mixing speed during emulsification with Gum Arabic (at 25 krpm), which is the reason why such experiments are not presented in this article.

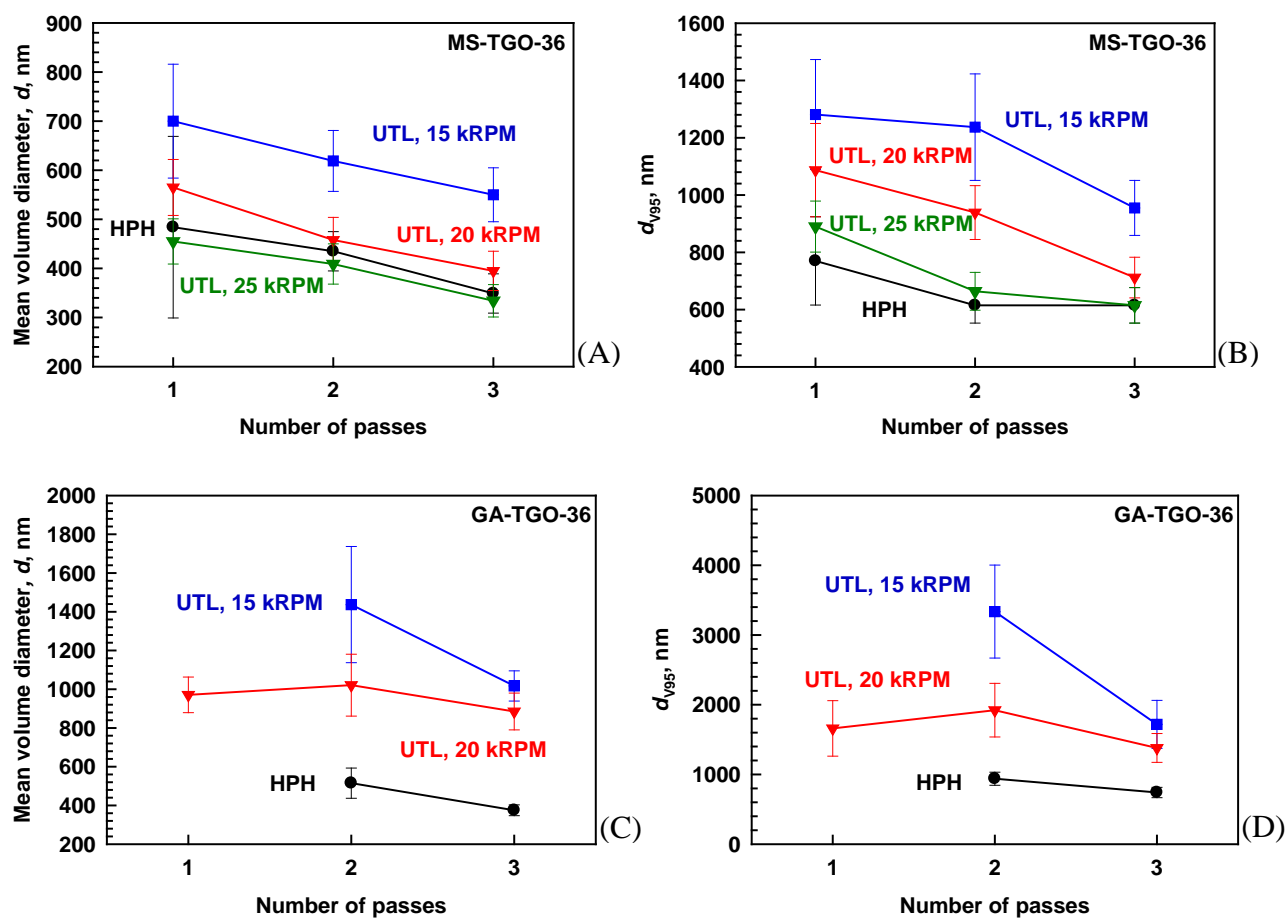

**Figure 3.** (A,C) Mean and (B,D) Maximum volume diameters of emulsion droplets. Emulsions in UTL contain (A, B) 17.9 wt% TGO-36 and 30.4 wt% MS, (C, D) 17.9 wt% TGO-36 and 30.4 wt% GA, whereas emulsions in HPH contain (A,B) 10 wt% TGO-36 and 15.6 wt% MS and (C, D) 10 wt% TGO-36 and 15.6 wt% GA. The rotational speeds of the UTL are indicated in the figures.

The reason for the different performance of MS and GA as emulsifiers could be related to: (1) lower viscosity of GA compared to MS, and (2) faster adsorption of MS compared to GA. Each of these factors could lead to pronounced coalescence upon emulsification for the GA-stabilized droplets, while the drop coalescence in MS solutions is expected to be lower. To clarify the relative importance of these factors, we first investigated the effect of the viscosity of the continuous phase on the drop size distribution in MS and GA emulsions. Because no effect of emulsifier concentration was observed on the interfacial tension in the studied concentration range, we increased the GA concentration from 30.4 to 35 wt% ($\eta_C$ = 120 mPa.s at 65 °C) to mimic the viscosity of the 30.4 wt% MS solution ($\eta_C$ = 110 mPa.s at 65 °C) which gave the smallest droplets in the UTL homogenizer. Afterwards, we used the 35 wt% GA solutions to prepare emulsions with 17.9 wt% TGO-36 at 20 krpm in the UTL. The obtained mean dropt sizes are compared in Figure 4 (maximum drop diameters and drop size distributions are shown in Figure B.3 in Supplementary material). The droplets prepared in the presence of Gum Arabic are much bigger than those obtained with MS as emulsifier. Therefore, the viscosity *per se* is not sufficient to explain the large size of the droplets in GA emulsions (compared to MS), but increasing it surely helps to decrease the drop size in these emulsions.

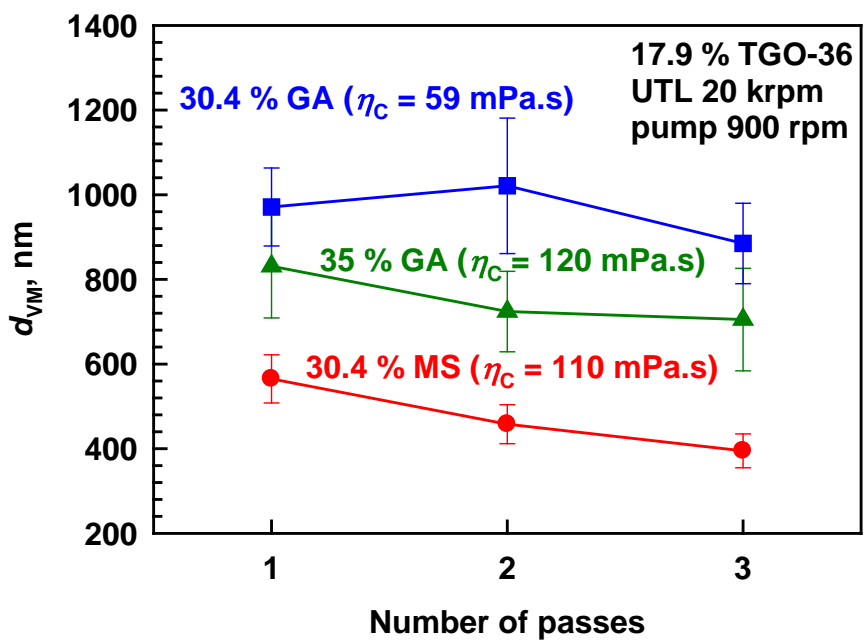

**Figure 4.** Mean volume diameters as a function of the number of passes through the UTL module at 20 krpm. Emulsions prepared with solutions of 30.4 wt% GA (blue squares), 35 wt% GA (green triangles) and 30.4 wt% MS (red circles).

To check whether the drop-drop coalescence is the main issue for emulsions prepared with GA, we performed experiments in which the concentration of GA was fixed and we varied the oil concentration in the emulsion. The obtained experimental results are shown in Figure 5. Increasing the oil mass fraction leads to a significant increase in the mean and maximal drop diameters in the formed GA emulsions (confirmed also by optical microscopy). Thus, the drop size in these emulsions is controlled by the emulsifier-to-oil ratio, due to a depletion of the surface active component in Gum Arabic in the process of emulsification (so-called "surfactant-poor regime"). Consequently, to prepare nanoemulsions with GA one should decrease the oil concentration (increasing the GA/oil ratio) or one should use an additional rapidly adsorbing co-surfactant.

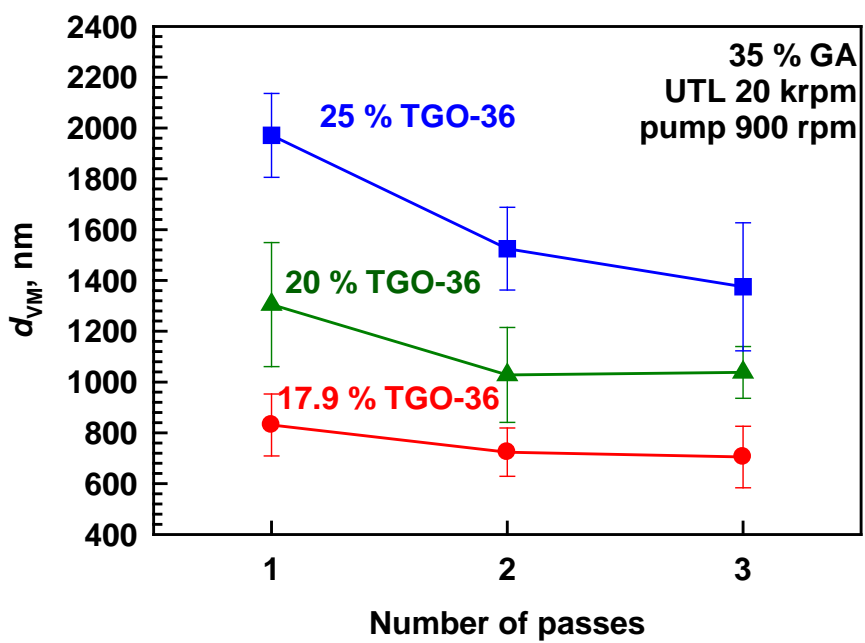

**Figure 5**. Mean volume diameters of droplets in emulsions with 17.9 % (red circles), 20 % (green triangles) and 25 % (blue squares) TGO-36, stabilized with 35 % GA. Emulsions prepared with UTL at 20 krpm.

### 3.2. Effect of oil drop viscosity on the formation of nanoemulsions.

We prepared emulsions of TGO-19 using HPH (10 wt% oil in the emulsions; 15.6 wt% emulsifier in the aqueous phase) and UTL (17.9 wt% oil in the emulsions; 30.4 wt% emulsifier). Two rotor speeds (20 and 25 krpm) were studied in the case of MS-emulsions, whereas the rotor speed was fixed to 20 krpm for the preparation of the emulsions with GA.

The mean drop diameters obtained in the emulsions with MS are presented in Figure 6. One sees that, when compared with emulsions of TGO-36, the mean and maximum drop diameters are smaller in emulsions of TGO-19 prepared with HPH, especially after 2 and 3 passes. With respect to the emulsions prepared with UTL, there is no significant effect of the oil viscosity on the drop size and similar results are obtained with both oils at 20 and 25 krpm, after 3 passes through the rotor-stator homogenizer. The direct comparison of the results obtained with HPH and UTL (presented in Figure B.6D in Supplementary material) shows that smaller drops are formed during emulsification with TGO-19 in HPH, compared to UTL. The observed effect of oil viscosity could be explained by the energy dissipation inside the deforming droplets during emulsification with HPH, whereas in the case of UTL there is enough time for droplet to deform. The latter is in a good agreement with the study of Tcholakova et al. (2011) where the authors show that the viscous turbulent regime realized in the rotor-stator type homogenizers is more efficient for oil drop breakup in the case of viscous oils.

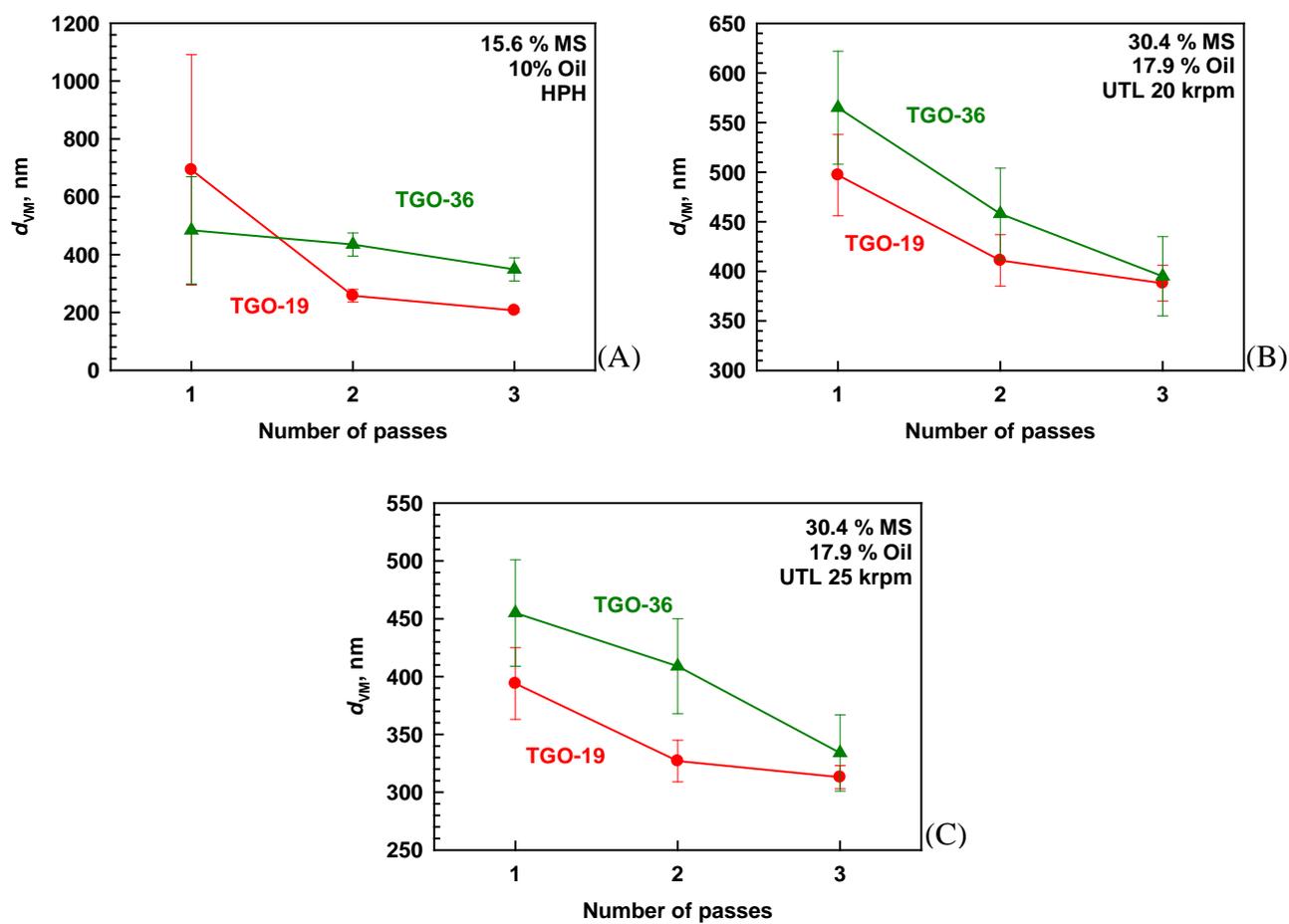

**Figure 6**. Mean volume diameters of droplets in emulsions stabilized with MS and prepared with (A) HPH at 5000 psi (10 % oil, 15.6 % MS) and with UTL (17.9 % oil, 30.4 % MS) at (B) 20 krpm and (C) 25 krpm. Red circles represent the drop size characteristics of emulsions with TGO-19 and green triangles – with TGO-36.

The drop size characteristics of TGO-36 and TGO-19-emulsions, stabilized with GA, are compared in Table 1. One sees that smaller droplets are again formed with TGO-19, due to its lower viscosity when HPH is used for emulsion preparation. On the other hand, the droplet sizes in emulsions of TGO-19 and TGO-36 formed in UTL are similar, as expected for emulsions with pronounced drop-drop coalescence. Note that during the 3$^{rd}$ pass through the UTL, the temperature of the emulsion of TGO-19 and GA increased up to 75 °C. The emulsifying ability of GA decreases significantly at temperatures higher than 70 °C. As a result, the droplet diameters in the emulsion obtained after 3 passes were bigger as compared to the emulsions from the previous passes, due to pronounced coalescence. Therefore, the results after 2 passes of the emulsions are compared in Table 1.

**Table 1.** Droplet size characteristics of emulsions formed in solutions of GA and prepared with UTL (17.9 % oil, 30.4 % GA) and with HPH at 5000 psi (10 % oil, 15.6 % GA).

| Homogenizer | Oil | $d_{VM}$, nm | $d_{V50}$, nm | $d_{V95}$, nm |
|---|---|---|---|---|
| UTL-20 krpm 2 passes, Pump: 500 rpm | TGO-19 | 851 ± 157 | 741 | 1480 |
| | TGO-36 | 796 ± 97 | 688 | 1380 |
| HPH 5000 psi, 3 passes | TGO-19 | 322 ± 12 | 295 | 620 |
| | TGO-36 | 375 ± 28 | 342 | 740 |

### 3.3. Effect of oil weight fraction on MS-stabilized emulsions

Two main series of emulsification experiments were performed to study the effect of oil concentration on drop size – emulsification at fixed emulsifier/oil ratio and emulsification at fixed emulsifier concentration in the aqueous phase. The results from these two series are presented and discussed separately below.

### 3.3.1. Effect of oil weight fraction on drop size distribution in emulsions with fixed emulsifier/oil ratio

For this series of experiments, we fixed at 1.4 the ratio emulsifier/oil. The concentrations of the other components (sodium benzoate and citric acid) were also recalculated depending on the oil weight fraction and emulsifier concentration (fixed emulsifier/solid component ratios). The TGO-36 concentration was varied from 16.1 to 20 wt%. These emulsions were prepared with UTL at 15 krpm and 20 krpm. In that way, we studied the effect of oil concentration at relatively constant interfacial tension, but at different viscosities of the aqueous phase. As was already discussed above, TGO-36-water interfacial tensions of the 26.9, 30.4 and 35 wt% MS solutions are very similar, but their viscosities are quite different – 53, 110 and 160 mPa.s, respectively (measured at 65 °C and 100 s$^{-1}$), see Figures 1 and 2. It might seem that the increase of oil concentration was small in this series of experiments. However, it was shown with GA solutions that even such small changes in the oil content could affect significantly the drops size when the emulsifier is unable to stabilize the drops during emulsification.

The mean drop diameters obtained after emulsification at 20 krpm are shown in Figure 7. One sees that drop sizes decrease with the increase of oil weight fraction due to the increase of the viscosity of the aqueous phase and of the emulsion. The mean drop diameter obtained at 20 krpm decreases from ~ 700 nm to ~400 nm and the maximum droplet diameter decreases from 1300 nm to 770 nm with the increase of oil concentration from 16.1 to 20 %.

Interestingly, there is no significant change in the drop size at oil concentration ≥ 17.9 wt%. It is rather unexpected that there is no significant difference in the drop size characteristics of the emulsions with 17.9 and 20 wt% TGO-36 (see drop size distributions shown in Figure B.9 in Supplementary material). To understand why the effect of the oil concentration disappears, we checked the temperatures of the emulsions measured after passing through the UTL module (see Table C.1A in Supplementary material). Very high temperature of 96 °C is reached during emulsification of 20 wt% oil emulsions, whereas the temperature of the emulsion with 17.9 wt% oil was ~70 °C. The temperature has a significant impact on both viscosities of the oil and of the aqueous phase. At 70 °C the solution of 30.4 % MS has viscosity ~78 mPa.s and the oil viscosity is ~12 mPa.s, whereas at 96 °C the viscosity of the solution of 35 % MS is ~110 mPa.s (extrapolated) and of the oil ~7 mPa.s (extrapolated). Therefore we expect that the decrease in the aqueous phase viscosity, caused by the significant increase in the temperature during emulsification, is the main reason for the observed similar drop sizes in emulsions with 17.9 and 20 % TGO-36.

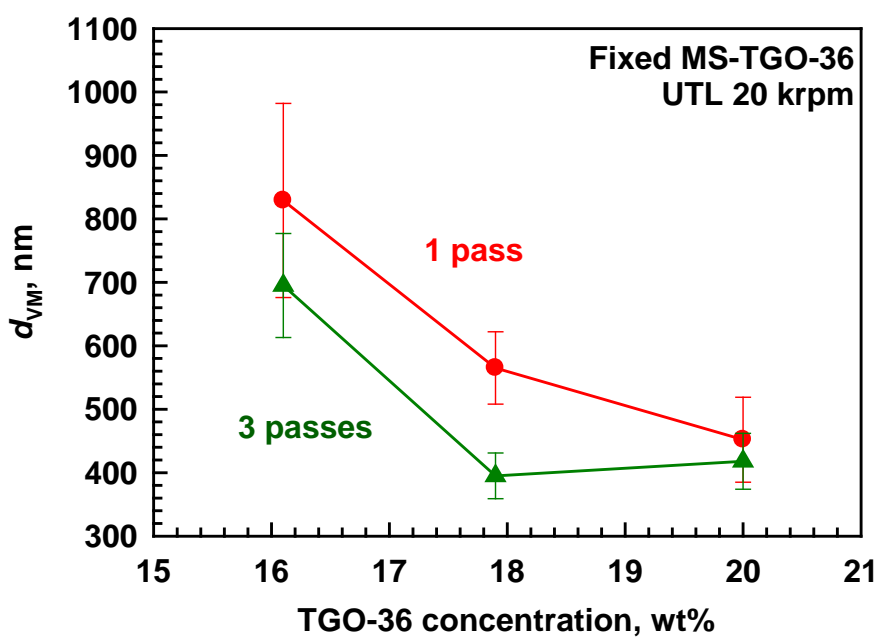

**Figure 7**. Mean volume diameters as a function of TGO-36 concentration. Emulsions stabilized with MS at fixed MS- TGO-36 ratio and prepared after 1 pass (red circles) and 3 passes (green triangles) through the UTL at 20 krpm.

## 3.3.2. Effect of oil weight fraction on drop size distribution in emulsions with fixed concentration of emulsifier in the aqueous phase.

In this part of the study we used 30.4 wt% solution of MS for emulsion preparation. TGO-36 and TGO-19 were used as dispersed phases. Three oil concentrations were studied – 17.9, 20 and 25 wt%. Emulsions were prepared with UTL at 20 krpm. To decrease the temperature during emulsification, we used a thermostat for the experiments with emulsions of 20 and 25 % oil. The obtained results are shown in Figure 8. One can see that the increase of oil weight fraction has small effect on the mean and maximum drop diameters for TGO-19 and practically no effect for TGO-36. The mean and maximum drop diameters in TGO-36-emulsion formed after 3 passes are 385 nm and 700 nm respectively, and in TGO-19-emulsion – 330 and 600 nm, i.e. smaller droplets are formed with the less viscous oil. The temperatures of the emulsions obtained for the three oil concentrations studied are similar (see Table C.1 in Supplementary material) which means that there is no significant change in the viscosity of the MS solution during emulsification. This result indicates that the main factor determining the drop size distribution is the viscosity of the aqueous and oil phases.

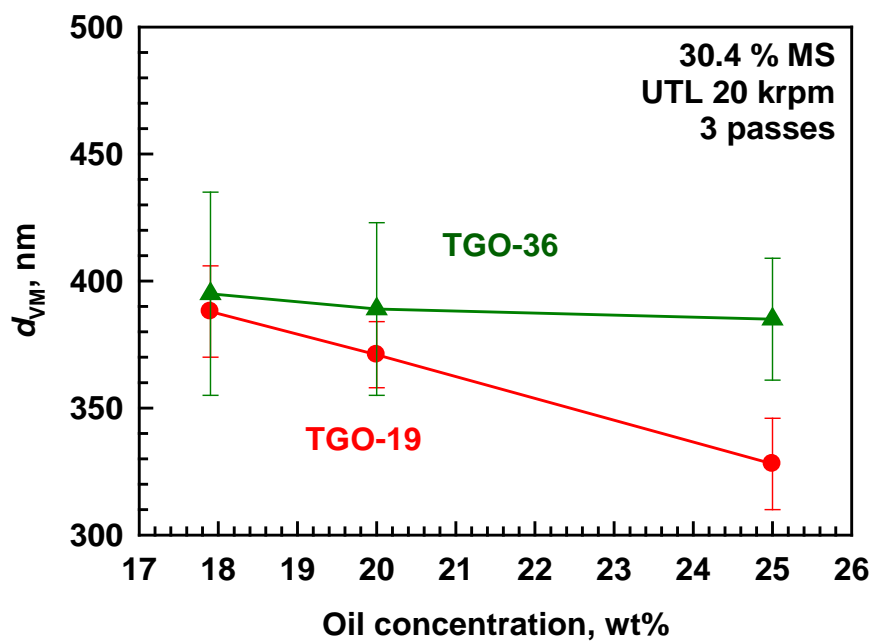

**Figure 8**. Mean volume diameters of droplets in emulsions with MS. All experimental conditions are illustrated in the figures.

## 4. Data interpretation and description of the results obtained with rotor-stator homogenizer.

### 4.1. Theoretical background: emulsification in viscous turbulent flow.

The classical studies of the emulsification process in turbulent flow, performed by Kolmogorov (1949) and Hinze (1955), showed that two qualitatively different hydrodynamic regimes of drop breakup should be distinguished – drop breakup in inertial regime of emulsification and drop breakup in viscous regime of emulsification. In the inertial regime, droplet deformation occurs under the action of pressure fluctuations, created by the irregular fluid velocity. The viscosity of the external phase is of secondary importance in this regime, because the inertial stress deforming the droplets is of inertial (i.e. non-viscous) origin. In contrast, drop deformation and breakup in the viscous regime is determined by regular local flows (shear, elongational, etc.) and by the related viscous stresses, which are created by velocity gradients. In this regime, the viscosity of the continuous phase is an important factor for the efficiency of drop breakup.

In the current study we work with very viscous solutions and small droplets, so our focus is mainly on the viscous turbulent regime of emulsification. In such systems, the drop breakup occurs under the action of viscous stress, $\tau_C$, inside the smallest turbulent eddies of the continuous phase which have a characteristic size, $\lambda_0 \approx \varepsilon^{-1/4} \eta_C^{3/4} \rho_C^{-3/4}$ (Hinze, 1955; Kolmogorov, 1949). The maximum stable drop diameter in this regime, $d_{KV}$, can be estimated by comparing $\tau_C$:

$$\tau_C = \eta_C \frac{dU_{\lambda 0}}{dx} \sim \eta_C \frac{(\eta_C \varepsilon / \rho_C)^{1/4}}{\lambda_0} \sim (\varepsilon \rho_C \eta_C)^{1/2} \quad (1)$$

to the drop capillary pressure, $P_C = 4\sigma/d$. In eqn. (1), the characteristic velocity inside the smallest eddies, $U_{\lambda 0}$, is found by balancing the inertial and viscous stresses in the smallest eddies, $\rho_C (\varepsilon \lambda_0)^{2/3} \sim \eta_C U_{\lambda 0}/\lambda_0$, which leads to $U_{\lambda 0} \sim (\eta_C \varepsilon / \rho_C)^{1/4}$. From $P_C$ and eqn. (1), the maximum droplet size, $d_{KV}$, is estimated in the viscous regime of emulsification as (Cristini, Blawzdziewicz, Loewenberg & Collins, 2003; Hinze, 1955; Kolmogorov, 1949):

$$d_{KV} = A_2 \varepsilon^{-1/2} \eta_C^{-1/2} \rho_C^{-1/2} \sigma \quad (2)$$

where $A_2$ is a numerical constant which depends on the viscosity ratio, $\eta_D/\eta_C$. As seen from eqn. (2), the droplet size depends on the viscosity of the continuous phase, $\eta_C$, in this regime of emulsification.

The main experimental results obtained with emulsions prepared in the viscous turbulent regime, clarified that often smaller droplets are obtained in this regime of emulsification (as compared to the inertial one) at similar all other conditions. This viscous regime was found to be particularly appropriate for successful emulsification of very viscous oils with $\eta_D$ up to 10 000 mPa.s (Tcholakova et al., 2011). The transition from inertial to the viscous turbulent regime of emulsification can be realized by a moderate increase of the viscosity of the aqueous phase ($\eta_C > 3$ mPa.s in the studied systems) and/or by an increase of the oil volume fraction, $\Phi > 0.6$. When the viscous regime is achieved by increasing $\Phi$ above 0.6, the polydispersity of the obtained concentrated emulsions of viscous silicone oils became significantly lower than that of the emulsions obtained at lower volume fraction. The observed significant decrease of the drop polydispersity indicated a change in the mode of droplet breakup (due to the specific microstructure of the concentrated emulsions), leading to the formation of more uniform droplets (Tcholakova et al., 2011).

As explained above, the steady-state drop size obtained in the viscous regime of emulsification could be estimated using the Kolmogorov-Hinze approach which compares the viscous stress, $\tau_C$, inside the smallest turbulent eddies of the continuous phase to the droplet capillary pressure, $P_C$. This approach could be applied to non-Newtonian continuous phases as well (Vankova et al., 2007). The experimentally measured droplet diameter, $d_{V95}$, was successfully described by a theoretical expression for the maximum stable droplet diameter, which takes into account the non-Newtonian rheological behavior of the concentrated emulsions of viscous oils (Vankova et al., 2007):

$$d_{V,PL} = A_7 \sigma c_\eta^{-1/(m+1)} (\varepsilon \rho_C)^{-m/(m+1)} \tag{3}$$

where $c_\eta$ is the emulsion consistency, $m$ is the power-law index ($0 < m \leq 1$), and $A_7$ is a numerical constant, which depends on the viscosity ratio $\eta_D/\eta_{EM}$ (and plays the role of the critical capillary number for droplet breakup in shear flow), see the work of Vankova et al. (2007) for more detailed explanations. For Newtonian continuous phase, Eq. (3) is simplified to Eq. (2).

### 4.2. Description of the experimental results.

To determine the value of $\varepsilon$ in the UTL-MagicLab homogenizer, we used the following expression (Calabrese, Chang & Dang, 1986; Calabrese, Wang & Bryner, 1986; Wang & Calabrese, 1986):

$$\varepsilon = b_1 N^3 r^2 \tag{4}$$

where $N$ is the rotation speed in revolutions per second and $r$ is the rotor radius in meters. The numerical constant $b_1$ was found to vary between 1 and 70, depending on the specific geometry of the rotor-stator head (Calabrese et al., 1986; Coulaloglou & Tavlarides, 1977;

Rushton, Costich & Everett, 1950). In our previous study (Tcholakova et al., 2011) we determined $b_1 = 40 \pm 4$ for UTL. This value of $b_1$ is used in all estimates hereafter. According to eqn. (4), $\varepsilon$ is equal to $1.4\times10^5$, $3.3\times10^5$ and $6.5\times10^5$ m$^2$s$^{-3}$ at rotor speeds of 15000, 20000 and 25000 rpm, respectively. The calculated rate of energy dissipation for high pressure homogenizer is 4-order of magnitude higher – it is $\approx 3.03\times10^9$ m$^2$s$^{-3}$, calculated using the known equation $\varepsilon = pQ/\rho_C V_{disp}$ (Tcholakova et al., 2011). For our experimental conditions, the applied pressure is $p = 34.47$ MPa, the flow rate is $Q = 3.3$ ml/s; the mass density is $\rho_C = 1050$ kg/m$^3$, and the volume where the dissipation takes place is $3.75\times10^{-11}$ m$^3$. The comparison of the experimental data shown in Figure 3A shows that the drop size in emulsions formed with UTL at 25 krpm is very similar to the size in the emulsions formed with high pressure homogenizer. The latter comparison shows that the drop breakup in emulsions prepared with UTL is by $> 10^4$ times energetically more efficient as compared to HPH homogenization.

Another important equipment-specific parameter that needs to be determined is the so-called "global shear rate", $\dot{\gamma}$. For that purpose we used the following expression (Tcholakova et al., 2011):

$$\dot{\gamma} = 2\pi r N / l \tag{5}$$

Here, $l$ is the gap-width between the rotor and stator. For the UTL homogenizing element of MagicLab, we estimated $\dot{\gamma} \approx 1.2\times10^5$, $1.6\times10^5$ and $2\times10^5$ s$^{-1}$ at rotor speeds of 15000, 20000 and 25000 rpm, respectively. Very high shear rates are achieved in UTL, because of the small gap between the rotor and stator.

As discussed above, the emulsions in presence of Gum Arabic exhibited an increase of the drop diameters with the increase of the oil volume fraction (at GA = *const*), whereas emulsions with MS did not undergo any variations of the drop size. We explained the observed increase of the drop size to depletion of the GA emulsifier and in the next subsection we prove that this assumption agrees with the obtained experimental data.

The size of droplets during emulsification depends on the concentration of emulsifier and on the hydrodynamic conditions. When working in an excess concentration of emulsifier (the so called "emulsifier-rich regime"), the droplet size distribution is controlled primarily by the hydrodynamic conditions (Tcholakova, Denkov & Lips, 2008; Tcholakova, Denkov, Sidzhakova, Ivanov & Campbell, 2003), since the drop-drop coalescence is negligible. In contrast, there is significant drop-drop coalescence in the emulsifier-poor regime, due to the limited coverage of the drop surface by emulsifying agent. Therefore, the size of the droplets is mainly controlled by their coalescence in the latter regime.

## 4.2.1. Emulsification in presence of coalescence. Description of the results obtained with GA.

The larger drop diameters, measured at higher oil content when GA is used as emulsifier, could be due to the drop-drop coalescence during emulsification or/and to formation of aggregates due to bridging induced flocculation. To distinguish between these two possibilities, we performed optical microcopy observations of the emulsion droplets and revealed that there are no visible flocs inside the GA-stabilized emulsions, while bigger droplets were seen at higher oil concentration. Therefore, the main reason for the larger drop size at higher oil concentration in the GA-stabilized emulsions is due to drop-drop coalescence.

In our previous work, we proposed simple phenomenological model which described very well the data in the "surfactant-poor" region for emulsions stabilized by whey proteins + 150 mM NaCl (Tcholakova, Denkov & Lips, 2008; Tcholakova, Denkov, Sidzhakova, Ivanov & Campbell, 2003). The main assumption in this model is that the drops continue to coalesce upon emulsification until the emulsifier adsorption on the drop surface reaches a certain threshold value, $\Gamma_M$. An additional assumption can be made to simplify the equations, namely, that virtually all available emulsifier is adsorbed on the drop surfaces in the course of emulsification. These assumptions, applied to a mass balance of the dissolved emulsifier (assumed equal to the adsorbed emulsifier), lead to the following expression for the mean drop diameter in the surfactant-poor regime:

$$d_{VM} \approx \frac{6\Phi}{(1-\Phi)} \frac{\Gamma_M}{\alpha C_{INI}} \qquad (6)$$

Here $d_{VM}$ is the mean drop diameter by volume, $\Phi$ is the oil volume fraction and $C_{INI}$ is the initial emulsifier concentration in the aqueous phase. $\alpha$ here has the meaning of an activity coefficient, which is to account for the fraction of the surface active protein in Gum Arabic. The value of $\alpha$ is expected to be around 2.4 % for Acacia Senegal (Montenegro, 2012). From the normalized mean drop diameter, $d_{VM}(1-\Phi)/\Phi$, as a function of oil mass fraction, we determined $\Gamma_M = 4.6$ mg/m$^2$ when using the results shown in Figure 5 which were obtained after 3 passes through the homogenizer. The value of $\Gamma_M$ has previously been found to vary between 4 and 7 mg/m$^2$ which is in a very good agreement with the estimate obtained in the current study (Padala, Williams & Phillips, 2009; Garti & Reichman, 1993).

To verify the conclusion that the worse performance of GA, when compared to MS, is related to drop-drop coalescence during emulsification we performed additional experiments in which 1 wt % Tween 20 or 1 wt % Tween 80 was added to 31.5 wt % GA and the emulsification was performed with UTL module at 20 krpm. The addition of fast adsorbing Tween 20 and Tween 80 led to significant decrease in the size of the formed drops – the drop size distribution of these emulsions became very similar to that in MS-stabilized emulsions even after 1 pass through the homogenizer, see Figure B.5 in Supplementary material. These results clearly show that the main problem with GA is the drop-drop coalescence during emulsification which can be overcome by introducing rapidly adsorbing surfactants.

**4.2.2. Emulsification in absence of coalescence: description of the results obtained with MS.**

As discussed above, there is no noticeable change in the drop size in MS-emulsions when the oil weight fraction is increased. Therefore, we assume that the drop coalescence is suppressed and the droplet size distributions are determined by the drop breakup mainly. In absence of coalescence, the theoretical model for emulsification in viscous turbulent regime could be used for data interpretation, however, one should know the hydrodynamic conditions (flow type, shear rate, rate of energy dissipation etc.) during emulsification.

As seen from the data shown in Table C.1 in Supplementary material, the temperature increases significantly during emulsification and thus can affect the rheological properties of the used dispersions. To account for the temperature effect on the viscosities of aqueous and oil phases, we fit the rheological data by using Andrade equation, $\lg \eta = A + B/T$, and we determined the values of $A$ and $B$ (see Figure D.1 and Table D.1 in Supplementary material). Then, we calculated the viscosity of the solutions and oils at the temperature of the emulsions after passing through UTL.

Before starting with the data analysis, one should determine the appropriate viscosity which will be used in the theoretical expressions. Figure D.2 in Supplementary material shows the comparison of the viscosity of the emulsions, stabilized by MS, and the viscosity of the respective aqueous phase. One sees that the presence of oil droplets leads to non-Newtonian rheological properties of the formed emulsions which affects the conditions during emulsification. It is more appropriate to use emulsion viscosity, $\eta_{EM}$, instead of the viscosity of the aqueous phase, as a control parameter, since $\eta_{EM}$ is much higher. Determining the emulsion viscosity, $\eta_{EM}$, is not a trivial task since, because one should account for the dependence of $\eta_{EM}$ on oil volume fraction, $\Phi$. For shear-thinning emulsions like those shown in Figure D.2 in Supplementary material, $\eta_{EM}$ should be taken at the appropriate shear rate which characterizes the emulsification flow. To determine the dependence of $\eta_{EM}$ on $\Phi$, we used the cell model of Yaron and Gal-Or (1972):

$$\eta_{EM} = \eta_C \left[ 1 + I\left(\Phi^{1/3}, p\right) \Phi \right] \qquad (7)$$

where $\Phi = m_{Oil} \Big/ \left[ \rho_{Oil} \left( \dfrac{m_{Oil}}{\rho_{Oil}} + \dfrac{100 - m_{Oil}}{\rho_C} \right) \right]$ is the oil volume fraction, $p = \eta_D / \eta_C$ is the viscosity ratio and $I(\Phi^{1/3}, p)$ is the following expression:

$$I(\Phi^{1/3}, p) = \dfrac{5.5\left[4\Phi^{7/3} + 10 - 84\Phi^{2/3}/11 + 4(1 - \Phi^{7/3})/p\right]}{10(1 - \Phi^{10/3}) - 25\Phi(1 - \Phi^{4/3}) + 10(1 - \Phi)(1 - \Phi^{7/3})/p} \qquad (8)$$

It was shown that the viscosity of Newtonian emulsions (non-shear-thinning emulsions) can be described relatively well by this model (Pal, 2000; Yaron & Gal-Or, 1972). Note that the temperature effect on the viscosities of the aqueous and oil phase is accounted in the calculation of $\eta_{EM}$ in this approach.

Additionally, we fit the curves in Figure D.2 in Supplementary material with the well-known Herschel-Bulkley equation and we determined the three parameters: yield stress, $\tau_0$; consistency, $K$ and power low index, $n$.

$$\tau = \tau_0 + K\dot{\gamma}^n \qquad (9)$$

Where $\tau$ is the total shear stress and $\dot{\gamma}$ is the shear rate. Then, we calculated the total shear stress and the apparent emulsion viscosity at global shear rate $1.6 \times 10^5$ s$^{-1}$, which corresponds to 20 000 rpm rotor speed in UTL. The comparison of the emulsion viscosities, calculated by eqns. (7-9), and the viscosity measured at 1000 s$^{-1}$ is given in Figure D.3 in Supplementary material. One sees that $\eta_{EM}$ obtained by Yaron's equation is close to the viscosities estimated for the global shear rate of $1.6 \times 10^5$ s$^{-1}$.

As a next step, we checked whether our experimental data for the drop size could be described by using the theoretical expression for maximal drop diameter in the viscous turbulent regime of emulsification:

$$d \approx \dfrac{\sigma}{(\varepsilon \eta_{EM} \rho_{EM})^{1/2}} k \qquad (10)$$

where $k$ is a correlation coefficient. As shown in Figure 9 for $d_{VM}$ and Figure D.4 in Supplementary material for $d_{V50}$ and $d_{V95}$, there is a good agreement between the experimental results and the theoretical predictions for both oils, TGO-19 and TGO-36. As far as these two oils have different viscosities, they have different values of $k$ for calculation of $d_{VM}$ which are 0.5 for TGO-19 and 0.63 for TGO-36. For the calculation of $d_{VMAX}$, the values of $k$ are close to 1 (0.9 for TGO-19 and 1.14 for TGO-36). Note that these slightly different values of $k$ are related to the different viscosity ratios for the aqueous and oil phases. Further experiments with oils having different viscosities are required to determine the full dependence $k = f(\eta_D/\eta_C)$.

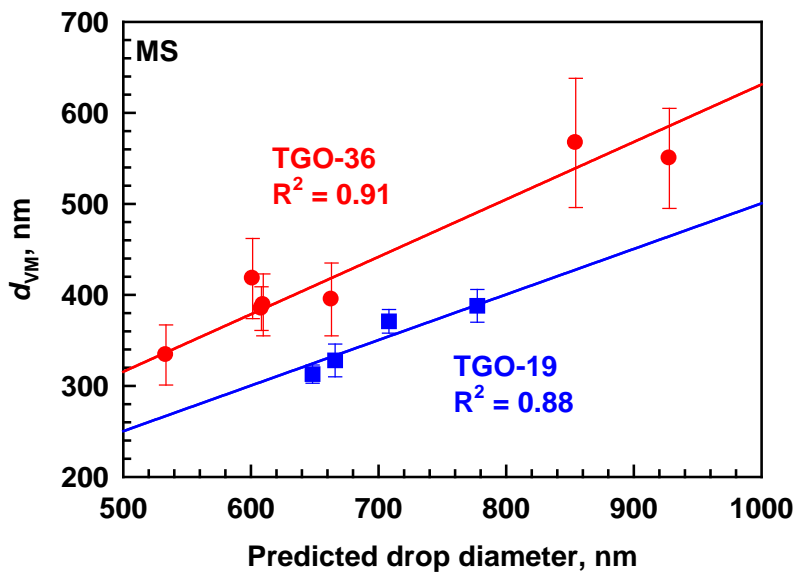

**Figure 9.** Correlation plot between the experimentally measured values of the mean drop diameter, $d_{VM}$, and the theoretically calculated values of $d$ using eqn. (10). Emulsions formed from MS solutions after 3 passes through the UTL at rotor speed of 15000, 20000 and 25000 rpm. Red symbols represent the results obtained with TGO-36 and the blue symbols – with TGO-19.

## 5. Conclusions.

In this study we show that nanoemulsions with droplet diameter as small as in emulsions obtained with high pressure homogenizer (in the range between 100 and 500 nm) could be prepared via rotor-stator homogenization for relatively short emulsification time when modified starch (MS) is used as emulsifier. The main advantage of MS, as compared to the low-molecular-mass surfactants, is its ability to increase the viscosity of the continuous phase and thus to facilitate the drop breakup during emulsification even at moderate and high oil volume fraction.

When Gum Arabic is used as an emulsifier, drop diameters smaller than 1 µm are formed only when HPH is used for emulsification and for low oil content in the emulsions. The main reason for the larger drops observed in the emulsions stabilized with GA, compared to MS-emulsions, is the significant drop-drop coalescence in the GA-stabilized emulsions, especially at high oil content of the emulsions. Possible routes to achieve smaller droplets stabilized by gum Arabic could be through reduction of the oil content or by addition of co-emulsifiers.

Theoretical analysis of the obtained experimental results shows that the droplet size in GA emulsions is strongly affected by the drop-drop coalescence during emulsification, whereas no such effect is observed in MS-stabilized emulsions. The experimental results for GA can be described under the assumption that there is a limited emulsifier adsorption on the drop surface which leads to drop coalescence, especially at higher oil volume fractions. The experimental results for the emulsions prepared with MS solutions are described well by the theoretical expression for emulsification in turbulent viscous regime, after proper account for the viscosities of the studied emulsions, including the effects of the temperature rise during emulsification.

## ACKNOWLEDGEMENT

The authors acknowledge the financial support from PepsiCo Global Beverage R&D, Hawthorne, New York, USA. This work was partially supported by the Bulgarian Ministry of Education and Science under the National Research Programme "Healthy Foods for a Strong Bio-Economy and Quality of Life" approved by DCM # 577/17.08.2018.

**Authors contributions:**

B.A. suggested studying the effect of GA and MS on the emulsification in HPH and rotor-stator homogenizer; S.T., N.D. and B.A. designed the experiments; D.G. and I.L performed the experiments and summarized the results; S.T. analyzed the results and proposed the theoretical description of the experimental results; D.G. wrote the first draft of the manuscript; S.T. prepared the final version of the manuscript. N.D. and B.A. edited the text.

## **Graphical abstract**

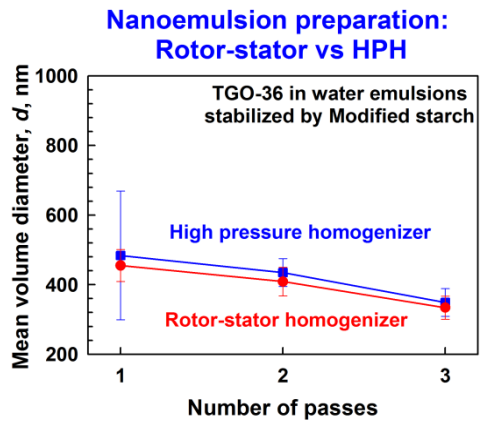
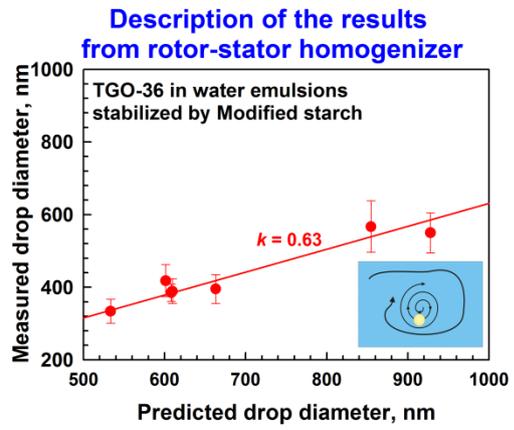

# Supplementary materials

## Food grade nanoemulsions preparation
## by rotor-stator homogenization

Dilek Gazolu-Rusanova[1], Ivan Lesov[1],

Slavka Tcholakova[1,*], Nikolai Denkov[1], Badreddine Ahtchi[2]

[1]*Department of Chemical and Pharmaceutical Engineering, Faculty of Chemistry and Pharmacy, Sofia University, 1164 Sofia, Bulgaria*
[2]*PepsiCo Global R&D, 3 Skyline Drive, Hawthorne, NY 10532, USA*

***Corresponding author**:
Prof. Slavka Tcholakova
Department of Chemical and Pharmaceutical Engineering
Faculty of Chemistry and Pharmacy, Sofia University
1 James Bourchier Ave., 1164 Sofia
Bulgaria

Phone: (+359-2) 962 5310
Fax:     (+359-2) 962 5643
E-mail: SC@LCPE.UNI-SOFIA.BG

**A. Viscosities, densities and oil-water interfacial tensions of the MS and GA solutions.**

**Table A.1**. Parameters of the linear fit of the data for the mass density, ρ, of the studied solutions and oils *vs.* temperature *T*.

(A) Aqueous solutions

| $\rho = a.T + b$ | MS concentration, wt% | | | | GA concentration, wt% | | |
|---|---|---|---|---|---|---|---|
| | 15.6 | 26.9 | 30.4 | 35 | 15.6 | 30.4 | 35 |
| *a* | -0.00038 | -0.00042 | -0.00046 | -0.00053 | -0.00039 | -0.00048 | -0.00048 |
| *b* | 1.066 | 1.116 | 1.132 | 1.157 | 1.069 | 1.138 | 1.162 |

(B) Oils

| $\rho = a.T + b$ | TGO-19 | TGO-36 |
|---|---|---|
| *a* | -0.0007 | -0.0007 |
| *b* | 0.9605 | 0.9372 |

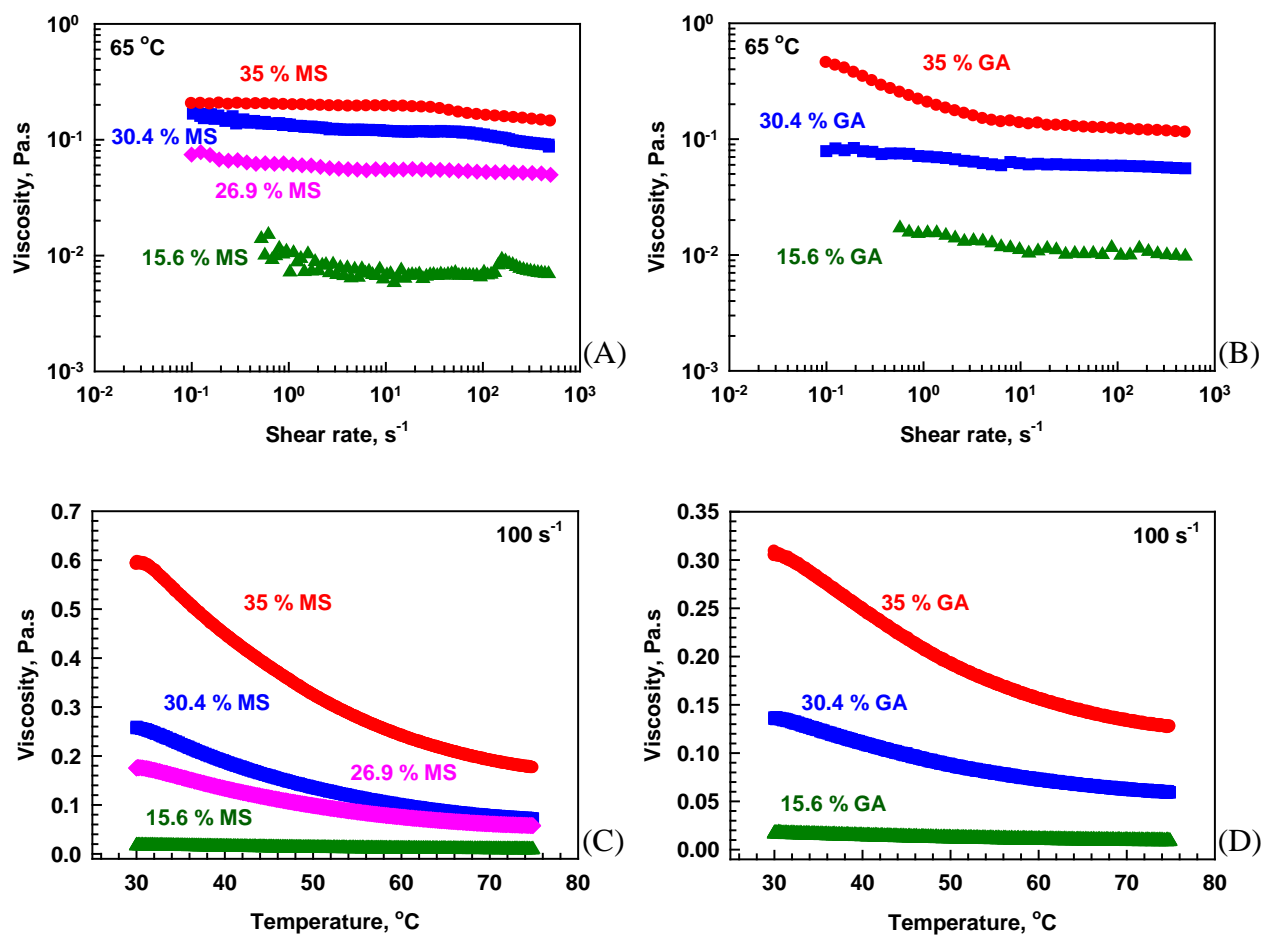

**Figure A.1.** Viscosity of the solutions of (A, C) MS, and (B, D) GA. Figures (A, B) show the viscosity as a function of shear rate, measured at 65 °C, while figures (C, D) show the viscosity as a function of temperature, measured at 100 s$^{-1}$.

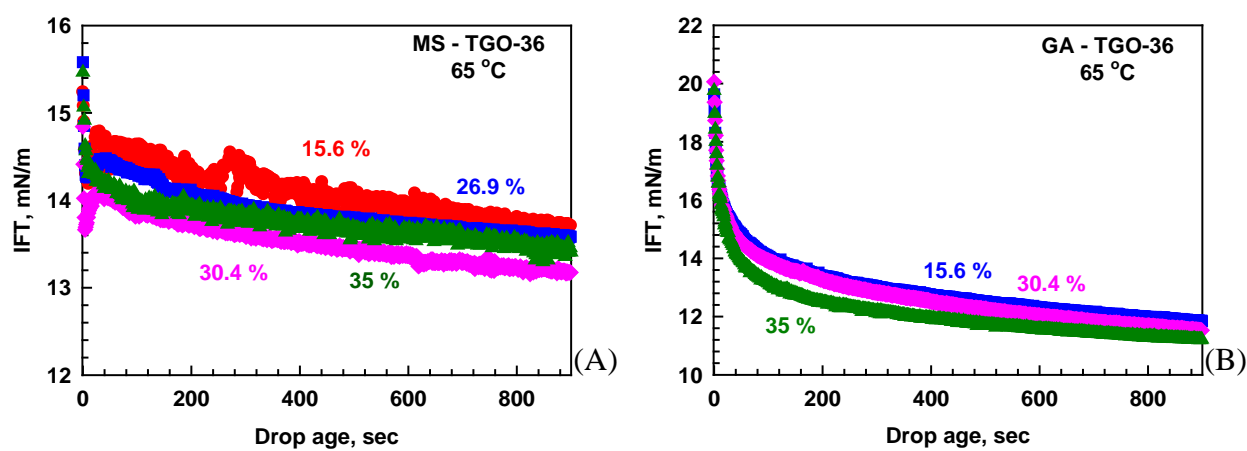

**Figure A.2.** Interfacial tension of (A) MS and (B) GA solutions, as a function of time at temperature $T = 65$ °C. Concentrations of biopolymers in wt% are color coded in the figures.

B. Drop diameters, $d_{V50}$ an

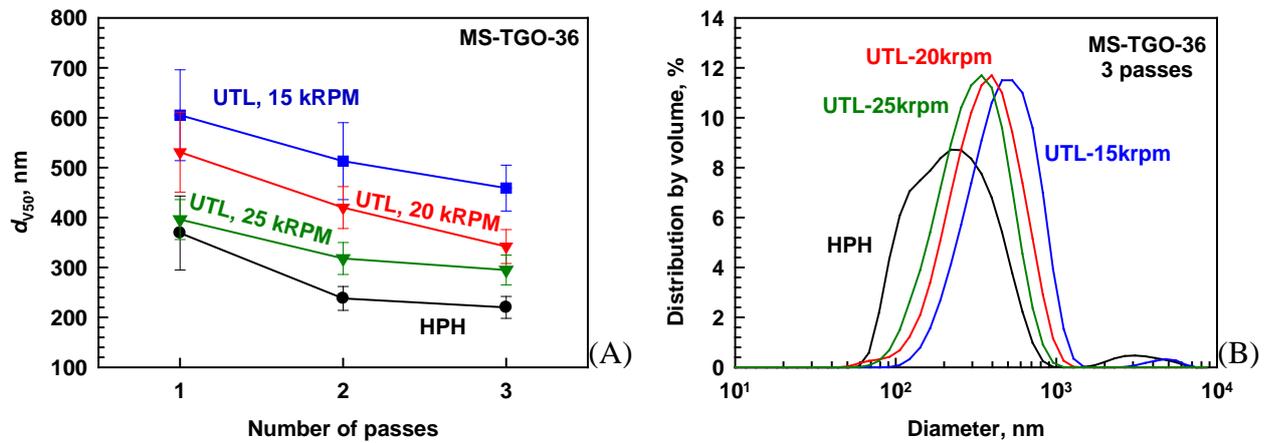

d $d_{V95}$, and drop size distributions in the emulsions obtained.

(A) TGO-36 emulsions

**Figure B.1.** (A) $d_{V50}$ as a function of the number of passes and (B) Drop size distributions by volume after 3 passes through the homogenizer. Emulsions obtained in UTL contain 17.9 wt% TGO-36 and 30.4 wt% MS, whereas the emulsions obtained in HPH contain 10 wt% TGO-36 and 15.6 wt% MS. The rotational speeds of the UTL are indicated in the figure.

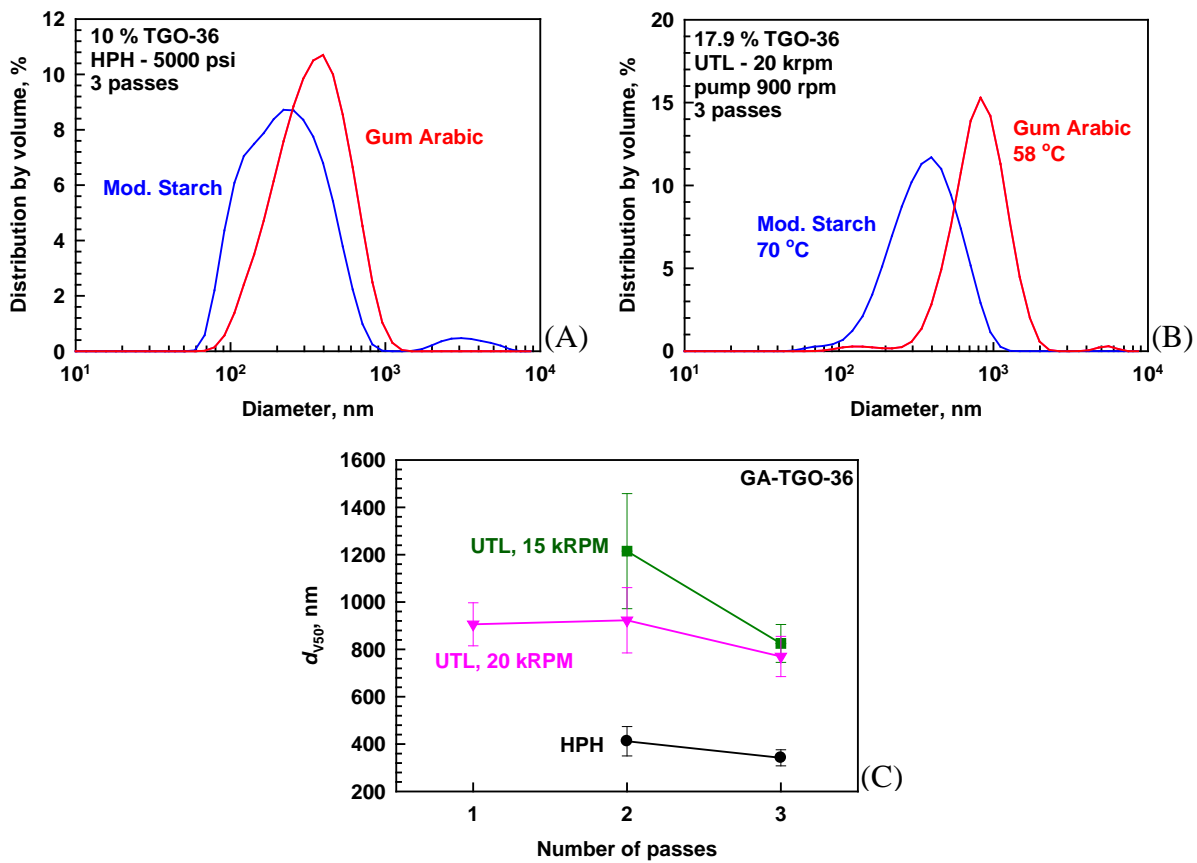

**Figure B.2.** Drop size distributions by volume in emulsions of MS (blue lines) and GA (red lines) prepared after 3 passes through (A) HPH – 10 wt% TGO-36 and 15.6 wt% emulsifier; and (B) UTL – 17.9 wt% TGO-36 and 30.4 wt% emulsifier. (C) $d_{V50}$ in emulsions of TGO-36 stabilized with GA, as a function of the number of passes through the HPH and UTL.

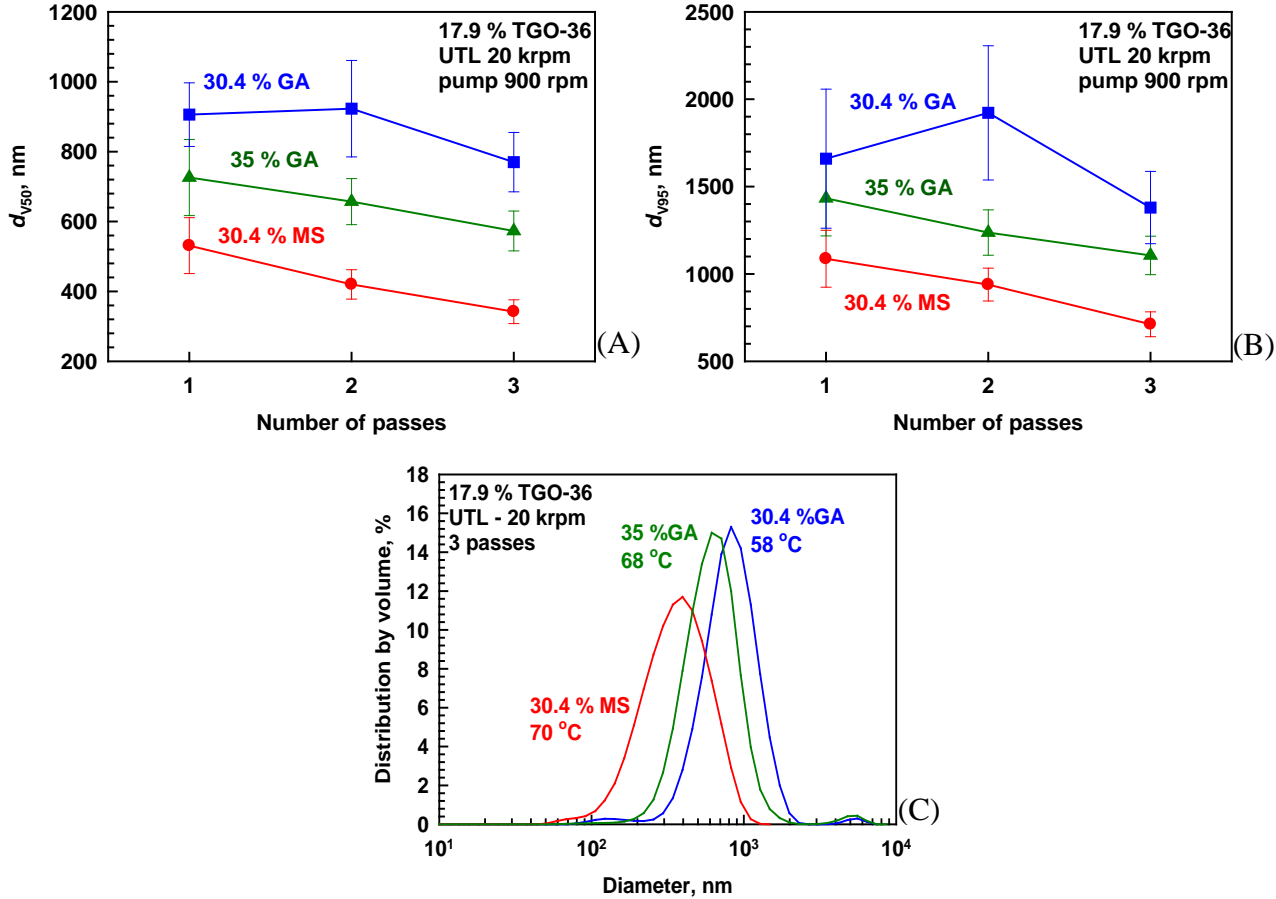

**Figure B.3.** (A) $d_{V50}$, (B) $d_{V95}$, and (C) Drop size distributions by volume in emulsions prepared with solutions of 35 wt% GA (green line), 30.4 wt% GA (blues line) and 30.4 wt% MS (red line).

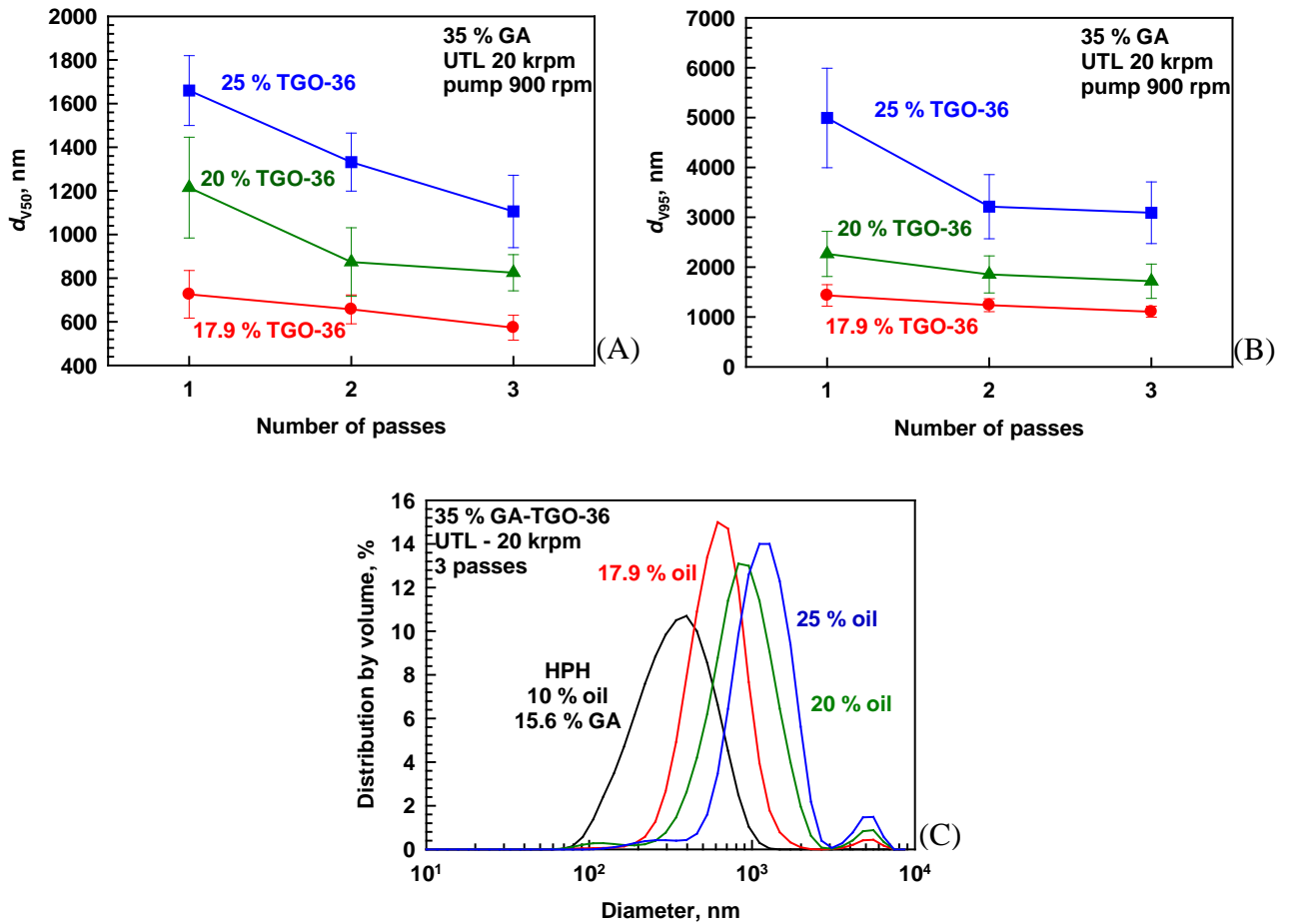

**Figure B.4.** (A) $d_{V50}$, (B) $d_{V95}$, and (C) Drop size distribution by volume in emulsions with 17.9 % (red lines), 20 % (green lines) and 25 % (blue lines) TGO-36, stabilized with 35 % GA. Emulsions prepared with UTL at 20 krpm. Black line represents the drop size distribution in emulsion prepared with HPH (10 % oil, 15.6 % GA).

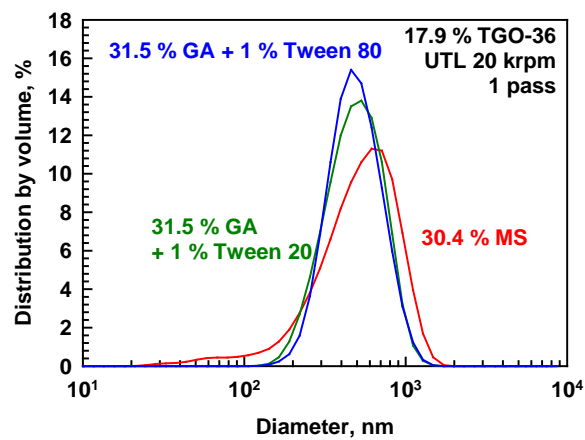

**Figure B.5.** Drop size distribution by volume in emulsions with 17.9 % TGO-36, stabilized by 30.4 % MS (red curve); 31.5 wt % GA+1 wt % Tween 80 (blue curve); or 31.5 wt % GA + 1 wt % Tween 20 (green curve). Emulsions are prepared by 1 pass through UTL at 20 krpm.

## C. Temperatures of the emulsions after passing through the homogenizer.

**Table C.1**. Temperatures of the emulsions with 10%, 17.9 %, 20 % and 25 % TGO-36 and TGO-19, stabilized with MS. Emulsions prepared with HPH at 5000 psi, or with UTL at 15 krpm, 20 krpm and 25 krpm rotor speeds.

**(A) TGO-36**

| Stabilizing agent | Oil concentration | HPH/UTL - RPM | Pump speed, rpm | Passes | T, °C |
|---|---|---|---|---|---|
| 15.6 % MS | 10 % TGO-36 | HPH – 5000 psi | - | 1, 2, 3 | 40-50 |
| 26.9 % MS | 16.1 % TGO-36 | 20,000 | 500 | 1 | 60 |
| | | | 500 | 2 | 66 |
| | | | 600 | 3 | 68 |
| 30.4 % MS | 17.86 % TGO-36 | 15,000 | 300 | 1 | 54 |
| | | | 300 | 2 | 60 |
| | | | 300 | 3 | 63 |
| | | 20,000 | 900 | 1 | 62 |
| | | | 900 | 2 | 68 |
| | | | 900 | 3 | 70 |
| | | 25,000 | 900 | 1 | 68 |
| | | | 900 | 2 | 75 |
| | | | 900 | 3 | 79 |
| | 20 % TGO-36 *Cooling with thermostat | 20,000 | 900 | 1 | 62 |
| | | | 900 | 2 | 65 |
| | | | 900 | 3 | 66 |
| | 25 % TGO-36 *Cooling with thermostat | 20,000 | 900 | 1 | 66 |
| | | | 900 | 2 | 70 |
| | | | 900 | 3 | 71 |

| Stabilizing agent | Oil concentration | HPH/UTL - RPM | Pump speed, rpm | Passes | T, °C |
|---|---|---|---|---|---|
| 35 % MS | 20 % TGO-36 | 20000 | 900 | 1 | 85 |
| | | | 900 | 2 | 92 |
| | | | 900 | 3 | 96 |

**(B) TGO-19**

| Stabilizing agent | Oil concentration | HPH/UTL - RPM | Pump speed, rpm | Passes | T, °C |
|---|---|---|---|---|---|
| 15.6 % MS | 10 % TGO-19 | HPH – 5000 psi | - | 1, 2, 3 | 40-50 |
| 30.4%MS | 17.86 % TGO-19 | 20000 | 500 | 1 | 62 |
| | | | 500 | 2 | 68 |
| | | | 500 | 3 | 72 |
| | | 25000 | 900 | 1 | 63 |
| | | | 900 | 2 | 70 |
| | | | 900 | 3 | 74 |
| | 20 % TGO-19 *Cooling with thermostat | 20000 | 900 | 1 | 60 |
| | | | 900 | 2 | 66 |
| | | | 900 | 3 | 67 |
| | 25 % TGO-19 *Cooling with thermostat | 20000 | 900 | 1 | 63 |
| | | | 900 | 2 | 69 |
| | | | 900 | 3 | 67 |

**Table C.2**. Temperature of the emulsions containing 10 %, 17.9 %, 20 % and 25 % TGO-36 or TGO-19, stabilized with GA. Emulsions prepared with HPH at 5000 psi, or with UTL at 15 krpm and 20 krpm rotor speed.

**(A) TGO-36**

| Stabilizing agent | Oil concentration | HPH/UTL - RPM | Pump speed, rpm | Passes | T, °C |
|---|---|---|---|---|---|
| 15.6 % GA | 10 % TGO-36 | HPH – 5000 psi | - | 1, 2, 3 | 30-35 |
| 30.4 % GA | 17.86 % TGO-36 | 15000 | 300 | 1 | 50 |
| | | 15000 | 300 | 2 | 55 |
| | | 15000 | 300 | 3 | 60 |
| | | 20000 | 900 | 1 | 51 |
| | | 20000 | 900 | 2 | 56 |
| | | 20000 | 900 | 3 | 58 |
| 35 % GA | 17.86 % TGO-36 | 20000 | 900 | 1 | 61 |
| | | 20000 | 900 | 2 | 68 |
| | | 20000 | 900 | 3 | 68 |
| | 20 % TGO-36 *Cooling with thermostat | 20000 | 900 | 1 | 58 |
| | | 20000 | 900 | 2 | 64 |
| | | 20000 | 900 | 3 | 56 |
| | 25 % TGO-36 *Cooling with thermostat | 20000 | 900 | 1 | 59 |
| | | 20000 | 900 | 2 | 68 |
| | | 20000 | 900 | 3 | 66 |

**(B) TGO-19**

| Stabilizing agent | Oil concentration | HPH/UTL - RPM | Pump speed, rpm | Passes | T, °C |
|---|---|---|---|---|---|
| 15.6 % GA | 10 % TGO-19 | HPH – 5000 psi | - | 1, 2, 3 | 30-35 |
| 30.4 % GA | 17.86% TGO-19 | 20000 | 500 | 1 | 60-54 |
| | | | 500 | 2 | 70 |
| | | | 500 | 3 | 75 |

## D. Interpolation of the experimental data.

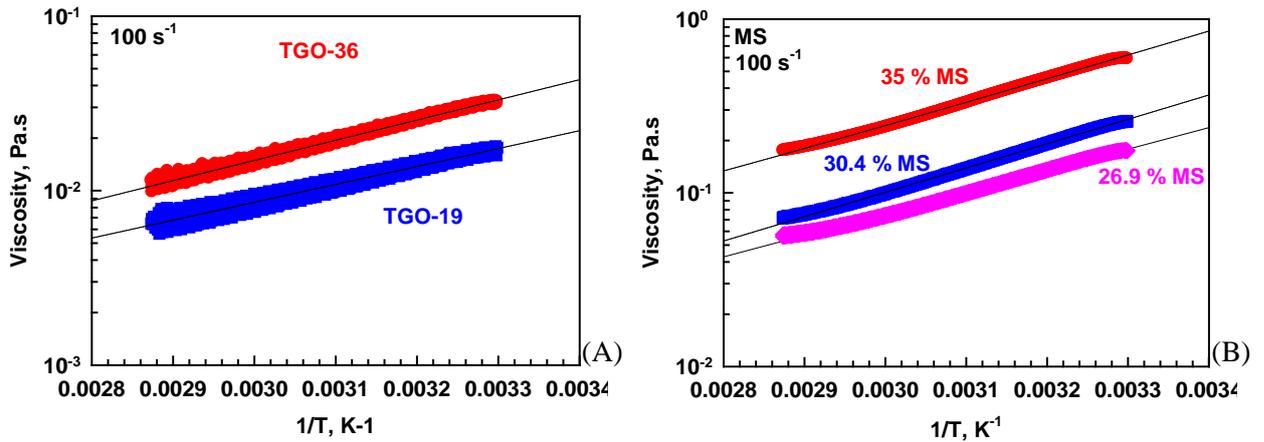

**Figure D.1**. Viscosity, η, of (A) used oils and (B) aqueous solutions with different concentrations of MS, as a function of the inverse temperature, $1/T$. The points are experimental data and the black curves are the best fits according to the equation $\lg \eta = A + B/T$.

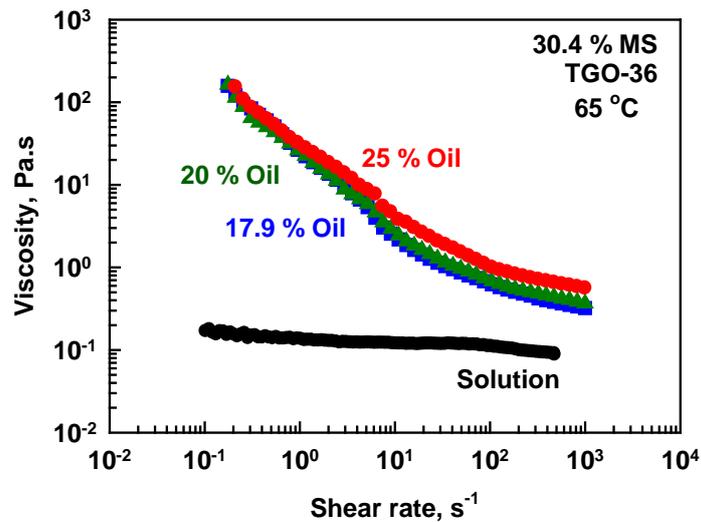

**Figure D.2**. Viscosity of MS-emulsions with 17.9 % (blue squares), 20 % (green triangles), 25 % TGO-36 (red circles), and viscosity of 30.4 % MS solution (black circles), as a function of shear rate, when measured at 65 °C.

**Table D.1.** (A) Values of the constants *A* and *B*, as determined by the best fits to the rheological data shown in Figure D.1 by the equation: $\lg \eta = A + B/T$. (B) Oil and aqueous phase viscosities and the respective viscosity ratio, calculated for 60, 70, and 80 °C using the equation: $\lg \eta = A + B/T$

(A)

|   | Oil phases | | Aqueous phases | | |
|---|---|---|---|---|---|
|   | TGO-19 | TGO-36 | 26.9 % MS | 30.4 % MS | 35 % MS |
| A | -5.133 | -5.290 | -4.84 | -5.13 | -4.62 |
| B | 1022 | 1154 | 1240 | 1380 | 1336 |

(B)

| $T$, °C | Oil viscosity, mPa.s | | 30.4 % MS solution viscosity, mPa.s | Viscosity ratio, $\eta_D/\eta_C$ | |
|---|---|---|---|---|---|
|   | TGO-19 | TGO-36 |   | TGO-19 | TGO-36 |
| 60 | 8.6 | 14.9 | 102.2 | 0.084 | 0.145 |
| 70 | 7.0 | 11.8 | 77.4 | 0.090 | 0.152 |
| 80 | 5.8 | 9.5 | 59.5 | 0.097 | 0.159 |

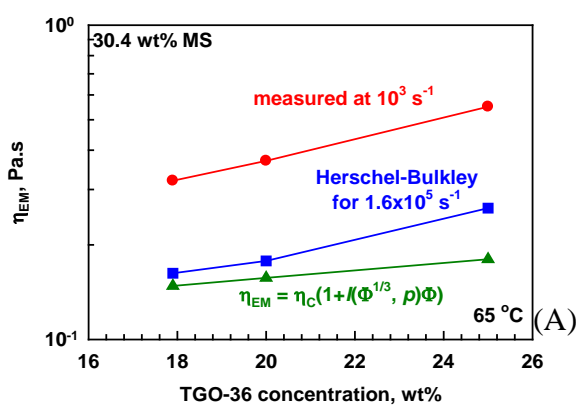
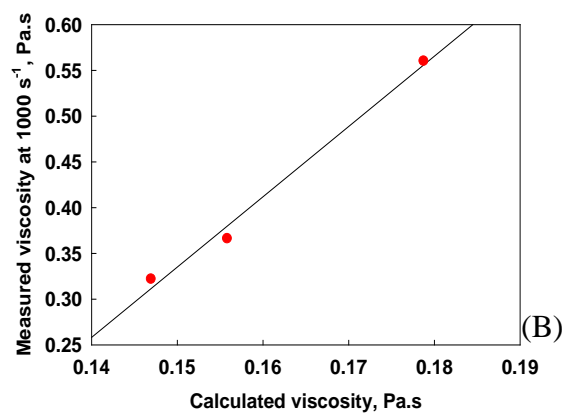

**Figure D.3.** Viscosity of MS-emulsions with different oil weight fractions: (A) measured at 1000 s$^{-1}$ (red circles), calculated by eqns. (7-8) (green triangles) and by eqn. (9) (blue squares); (B) measured at 1000 s$^{-1}$ (red circles) *vs.* calculated by eqns. (7-8). Temperature $T = $ 65 °C in all cases.

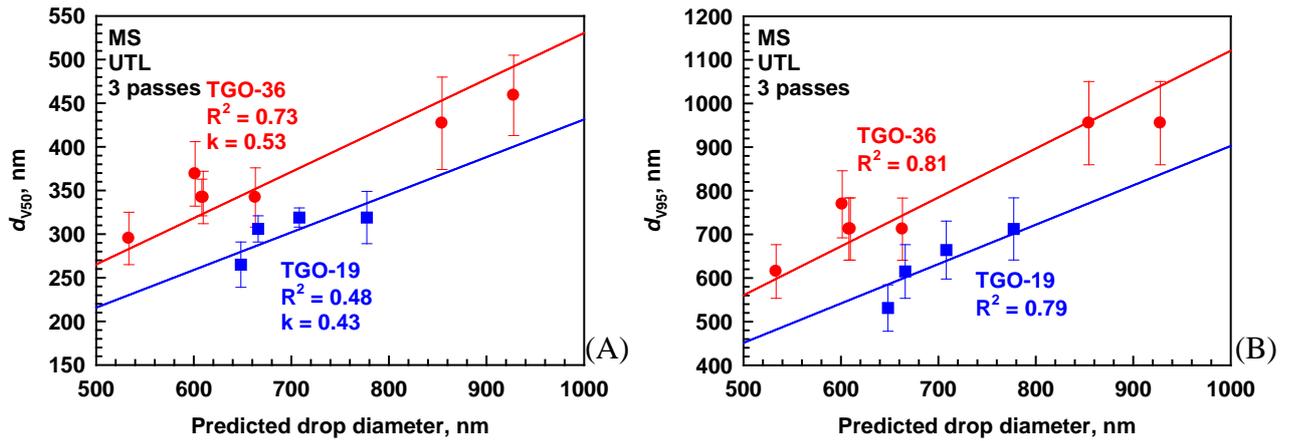

**Figure D.4.** Correlation plot between the experimentally measured values of (A) $d_{V50}$ and (B) $d_{V95}$, *vs.* the theoretically calculated values of $d$ by eqn. (10). Emulsions are formed in MS solutions after 3 passes through the UTL homogenizer at 15000, 20000 and 25000 rpm rotor speed. Red symbols represent the results obtained with oil TGO-36 and the blue symbols – with TGO-19.